\documentclass[12pt,preprint]{aastex}

\usepackage{caption}
\usepackage{graphics,epsf}
\usepackage{amsmath}                
\usepackage{amsfonts}               
\usepackage{amssymb}                
\usepackage{epsfig}                 
\usepackage{graphicx}
\usepackage{caption}
\usepackage{subcaption}
\usepackage{url}

\def \cm{~\rm{cm}}
\def \s{~\rm{s}}
\def \km{~\rm{km}}

\def \K{~\rm{K}}
\def \g{~\rm{g}}

\def \erg{~\rm{erg}}

\begin{document}

\title{Exploding Core-Collapse Supernovae by Jets-Driven Feedback Mechanism}

\author{Oded Papish\altaffilmark{1} and Noam Soker\altaffilmark{1}}

\altaffiltext{1}{Department of Physics, Technion -- Israel Institute of Technology, Haifa
32000, Israel; papish@physics.technion.ac.il; soker@physics.technion.ac.il}

\begin{abstract}
We study the flow structure in the jittering-jets explosion model of
core-collapse supernovae (CCSNe) using 2.5D hydrodynamical simulations and
find that some basic requirements for explosion are met by the flow.
In the jittering-jets model jets are launched by intermittent accretion disk
around the newly born neutron star and in stochastic directions. They
deposit their kinetic energy inside the collapsing core and induce explosion
by ejecting the outer core. The accretion and launching of jets is operated
by a feedback mechanism: when the jets manage to eject the core, the
accretion stops.
We find that even when the jets' directions are varied around the symmetry
axis they inflate hot bubbles that manage to expel gas in all directions. We
also find that although most of the ambient core gas is ejected outward,
sufficient mass to power the jets is accreted ($\sim 0.1 M_\odot$), mainly from the equatorial plane direction. This is compatible with the jittering jets explosion mechanism being a feedback mechanism.
\end{abstract}

\section{INTRODUCTION}
\label{sec:intro}
Baade and Zwicky \citep{Badde1934} in their seminal work were the first to suggest that supernovae (SNe) draw their energy from the gravitational collapse of a star into a neutron star (NS). After almost a century the processes by which part of this gravitational energy is channeled to explode the star remain controversial.

\cite{Colgate1966}  suggested that neutrinos emitted by the cooling newly formed NS play the major role in setting the explosion. Later \cite{Wilson1985} and \cite{bethe1985} develop this idea into the delayed-neutrino mechanism. In this model neutrinos emitted within a period of $\sim 1 \s$ after the bounce of the collapsed core heat material at the gain region ($r \approx 100-200 \km$) enough to expel the rest of the star.
In the last two decades sophisticated multidimensional hydrodynamical simulations with ever increasing capabilities  were  used
to study the delayed neutrino mechanism (e.g., \citealt{bethe1985,Burrows1985,Burrows1995,Fryer2002,Ott2008,Marek2009,Nordhaus2010,Kuroda2012,Hanke2012,Bruenn2013}).
This mechanism, although appealing, remains controversial. Highly sophisticated multidimensional simulations in the last several years did not manage to explode
the star with the observed kinetic energy. 
It seems that in most of the cases where explosion with an energy of $~10^{51} \erg$ was achieved, the explosion was driven by a continuous wind from the center.
{{{  {For example, we examine the results of Bruenn et al. (2013) and find that the increase of energy with time fits
an explosion by a continuous wind. More details of their simulations are required to fully capture the exploding agent.}  }}} {{{{Such winds were suggested to power supernovae explosions in the past (\citealt{Burrows1993,Burrows1995}) but where found to have limited contribution to the explosion for a more massive than $8.8 M_\odot$ starts.}}}}
In many cases the simulations have failed to even revive the shock of the infalling core material. In a study that scaled energy according to SN 1987A, it was found that even if neutrino explosion does work for some CCSNe, no explosions with kinetic energy of $>2 \times 10^{51} \erg$ are achieved \citep{Ugliano2012}. However, many CCSNe, e.g., some recent Type Ic SNe (\citealt{Roy2013,Takaki2013}) seem to have kinetic energy of
$ \ga 10^{52} \erg$.

Recent results have shown that in more realistic 3D numerical simulations explosions are even harder to achieve
{{{{  \citep{Janka2013, Couch2013,Takiwaki2013, Hanke2012, Hanke2013, couch2013arXiv}. However, \cite{Nordhaus2010} and \cite{Dolence2013}
obtained opposite results, i.e., it is easier to achieve shock revival in 3D simulations. }}}}
In a previous paper \citep{Papish2012b} we argued that the delayed-neutrino mechanism has a generic character that prevents it from exploding the star with the observed energy of $\sim 10^{51} \erg$. As written above, some of the problems of the delayed-neutrino mechanism can be lifted if there is a strong neutrino wind.

The lack of persisting success in exploding core collapse supernovae (CCSNe) with the delayed-neutrino mechanism and the diversity of CCSNe being discovered in recent years (e.g., \citealt{Arcavi2012}) show the importance of considering alternative models for  CCSN explosions.
Among the variety of alternative explosion mechanisms \citep{Janka2012} the magnetohydrodynamics class of models is one of the well studied
(e.g. \citealt{LeBlanc1970,  Meier1976, Bisnovatyi1976, Khokhlov1999, MacFadyen2001,Hoflich2001, Woosley2005, Burrows2007, Couch2009,Couch2011,Lazzati2011}).

In most of the MHD models the magnetic field is enhanced by a rapidly rotating core through the magnetorotational mechanism,  creating bipolar outflows (jets)
that are able to explode the star. Most stellar evolution models show that the core's rotation rate in most cases is too low for the magnetorotational mechanism
to be significant, making most of these models applicable for only special cases.
Recent observations (e.g. \citealt{Milisavljevic2013,Lopez2013}) show that jets might have a much more general role in CCSNe than that which is suggested by these models.

In \cite{Soker2010} and \cite{papish2011} we proposed a jets-based mechanism that might be present in all CCSNe explosions, and where the sources of angular momentum
are the convective regions in the core \citep{GilkisSoker2013} and/or instabilities in the shocked region of the collapsing core, e.g.,
the standing accretion shock instability (SASI),
{{{ {e.g., \cite{BlondinMezzacappa2007}, \cite{Fernandez2010}, and \cite{Rantsiouetal2011} who suggested that the source of the angular momentum of
pulsars is the spiral mode of the SASI.} }}}
Recent 3D numerical simulations show indeed that the SASI is well developed during the first second after core bounce (e.g., \citealt{Hanke2013}).
Another motivation to consider jet-driven explosion mechanisms is that jets might supply the site for the r-process \citep{Winteler2012,Papish2012b}.

In the present study we conduct 2.5D numerical simulations to study the explosion of the
core in the frame of the jittering-jet explosion mechanism \citep{papish2011}.
The main ingredients and assumptions of this mechanism are as follows.
 (1) We do not try to revive the stalled shock, to the contrary, our model
requires the material near the stalled-shock to fall inward and form an accretion
disk around the newly born NS or black hole (BH).
(2) We conjecture that due to envelope stochastic processes,  such as convection
and the SASI, segments of the post-shock accreted gas (inward to the stalled shock wave) possess local
angular momentum. When they accreted they form and accretion disk with rapidly varying axis direction.
In a recent study Gilkis \& Soker (2013) have shown that the convective
region in the oxygen burning shell, and in many cases the silicon-burning shell, will indeed lead  to stochastic angular momentum accretion
that is sufficient to form intermittent accretion disk around the newly born NS.
(3) We assume that the accretion disk launches two opposite jets. Due to the rapid change in
the disk's axis, the jets can be intermittent and their direction rapidly varying.
These are termed {\it jittering jets}.
(4) We analytically showed  in \cite{papish2011} that the jets penetrate the infalling gas up to a distance of ${\rm few} \times 1000 \km$,
i.e., beyond the stalled-shock. However, beyond ${\rm few} \times 1000 \km$ the jets cannot penetrate the gas
any more because of their jittering. The jittering jets don't have the time to drill holes through the ambient gas
before their direction changes; they are shocked before penetrating through the ambient gas.
This condition can be met if the jets' axis rapidly changes its direction. This process of depositing jets' energy
into the ambient medium to prevent further accretion is termed the {\it non-penetrating jet feedback mechanism} \citep{Soker2013}.
(5) The jets deposit their energy inside the collapsing star via shock waves, and form two hot bubbles,
that eventually merge and accelerate the rest of the star and lead to the explosion.
(6) The jets are launched only in the last phase of accretion onto the NS. For the required energy the jets must be
launched from the very inner region of the accretion disk. In this study we explore the flow structure described in points (4) and (5) above.

The paper is organized as follows. In section \ref{sec:numerical} we describe the numerical setup for the 2.5D simulations. In section \ref{sec:results} we describe our results, focusing on the general flow structure and the way the explosion starts in the core. In section \ref{sec:acc} we show that sufficient mass is accreted to close the  feedback cycle. In the present study we do not simulate the entire star. Our summary is in section \ref{sec:summary}.
\section{NUMERICAL SETUP}
\label{sec:numerical}
We use the 4th  version of the {\sc flash} gasdynamical numerical code  \citep{Fryxell2000}.
The widely used {\sc flash} code is a publicly available code for supersonic flow suitable for astrophysical applications.

The simulations are done using the Unsplit PPM solver of {\sc flash}. We use 2D axisymmetric cylindrical coordinates with an adaptive mesh refinement (AMR) grid; this is referred to as 2.5D type of simulations. The grid is divided into two regions. (1) A spherical inner region with a radius of about $1000 \km$. Inside this region we set the maximum refinement level to be proportional to angular size. This is up to a radius of $\sim 300 \km$ where the maximum refinement level is constant and with a resolution of about $1 \km$. (2) In the region outside $1000 \km$ we set a constant maximum refinement level which coincides with the maximum refinement level at the outer boundary of the inner region; the resolution in the outer region is $\sim 15 \km$. This separation into two different regions enables us to simulate both the region near the NS where we inject the jets and the large scale regions where the jets interact with the surrounding core material. The total size of the grid is $15,000 \times 30,000 \km$. We treat the spherical inner region of up to $75 \km$ from the center as a hole. This means that we are not simulating the NS itself, nor the assumed accretion disk. The boundary condition at the edge of the hole is inflow only, meaning the velocity cannot be positive in the radial direction unless we inject a jet at that specific zone.

We start the simulations after the core has collapsed and bounced back. The initial conditions are taken {{{ {for a $15 M_\odot$ model} }}} from the 1D simulations of
\cite{Liebend2005} at a time of $t \simeq 0.2 \s$ after bounce. We map their results into 2D, including the chemical composition.
We set outflow boundary conditions at the exteriors of the simulation's domain.

The equation of state (EOS) was chosen to be the Helmholtz EOS \citep{timmes2000} implemented by {\sc flash}. This EOS includes contributions from partial degenerate electrons and positrons,
radiation, and non degenerate ions. It is appropriate for our simulations as we do not simulate the most inner part where nuclear density is reached.
We added to the code neutrino cooling. The cooling rate is taken to be
\begin{equation}
Q_{\nu}^- = 5 \times 10^{30} \frac{T_{10}^{9}}{\rho_6} + 9\times 10^{23} X_{\rm nuc} T_{10}^6 \erg \g^{-1} \s^{-1},
\end{equation}
here $T_{10}$ is the temperature in $10^{10} \K$, $\rho_6$ is the density in $10^6 \g \cm^{-3}$, and $X_{\rm nuc}$ is the mass fraction of free nucleons.
The first term is the contribution of electron positron pair annihilation \citep{Itoh1996} and the second is due to electron/positron capture on free nucleons \citep{Qian1996}.
Gravity is treated using a point mass of $1.4 M_\odot$ at the center plus a contribution from the monopole term ($l=0$) in the simulation's domain.

We inject the jets during their activity inside the inner boundary with a velocity of $v_j = 10^{10} \cm \s^{-1}$ and a density $\rho_j$ derived from the required outflow rate of $\dot M_j = 4 \pi \delta r^2 \rho_j v_j $, here $\delta$ is the opening solid angle; $M_j$ is the total outflow mass in the two bipolar jets. 
{{{{There is a delicate matter of translating jets from 3D to 2.5D simulations. One cannot maintain both ram pressure and power from 3D to 2.5D as non-axial jets have the shape of a conical shell in 2.5D. In our simulations we keep the ram pressure constant by keeping the density and velocity as in 3D simulations.
To preserve energy and mass we activate the jets for shorter times such that the total energy and mass in each episode is the same as in 3D simulations.}
}}}

\begin{figure}[h!]
\begin{center}
\includegraphics[width=0.3\textwidth]{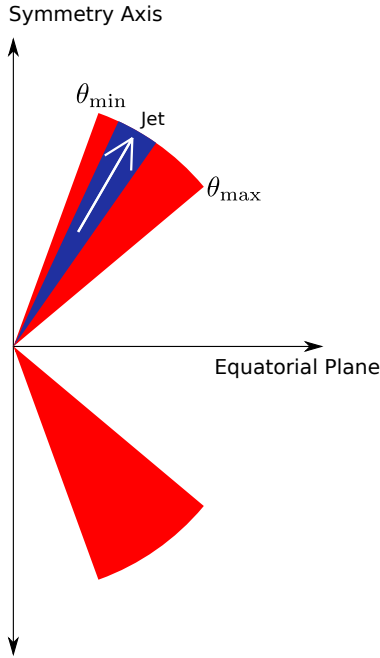}
\caption{The geometry of the simulations. The red region depicts the range of angles within which the jets are jittering.
The width of a jet-episode, which is actually a conical shell due to the 2.5D nature of the simulated grid, is shown with the blue
color. }
\label{fig:angles}
\end{center}
\end{figure}

The schematic jets' launching angles are depicted in Figure \ref{fig:angles}.
The total width of the jets is $\Delta \theta=2\alpha$.
We actually inject a conical shell due to the 2.5D nature of the grid.
In each pulse material is injected within a conical shell given by $\theta_{jc} \pm \alpha$,
where $\theta_{jc}$ is the center of the shell.
{{{ {Such a conical shell might actually be formed by a very rapid precession of the accretion disk, hence jets' launching direction,
around an axis; hence our term of 2.5D simulations.
We note that the jittering-jets explosion model requires no initial core angular momentum. However, the 2.5D simulations performed here are
limited to jittering around a preferred axis (the symmetry axis of the grid). A non-negligible amount of angular momentum in the pre-collapse core
can cause the `jittering' to take place around a preferred axis, e.g., forming a conical shape as simulated here.} }}}
The center of each jet's-shell $\theta_{jc}$ is varied within the range $\theta_{\rm min} - \theta_{\rm max}$.
The range of angles and the jittering prescription are varied between the different runs, as described in the next section.
\section{RESULTS}
\label{sec:results}
\subsection{The Global Flow Structure}
\label{subsec:global}

We first describe the results for a run we term 2s20-50, where the `2' stands for the total jets' activity duration,
and the the numbers `20-50' stand for the boundary angles of the jittering $\theta_{\rm min}=20^\circ$ and $\theta_{\rm max}=50^\circ$,
as drawn in Fig. \ref{fig:angles} and given in Table \ref{Tab:Table1}.
We start this run by simultaneously launching two jets, one at each half of the equatorial plane.
They are not exactly opposite to mimic the stochastic nature of the jittering jet model (see angles in Table \ref{Tab:Table1}).
The total jets' mass of all episodes combined and the launching velocity are the same for the three runs {with jets} studied here, with values of
$M_j = 0.026 M_\odot$ and $v_{ j} = 100,000 \km \s^{-1}$, respectively.
The initial total energy of the jets is
$E_{j}=E_{j,{\rm kin}} + E_{j,{\rm bind}}\simeq 2 \times 10^{51} \erg$, where $E_{j,{\rm kin}}$ is the kinetic energy of the jets and
$E_{j,{\rm bind}}$ is the gravitational binding energy of the jets. The initial thermal energy is negligible.
The mass and energy are distributed equally among the 10 jets'-launching episodes.
\begin{table}[h]
        \centering
    \begin{tabular}{lccccp{4cm} p{5cm}}
    \hline \hline
    Model   & $\dot M_j$  & $t_j$  & $\theta_{\rm min}$ & $\theta_{\rm max}$  & $\theta_{u}$           & $\theta_{d}$\\
    \hline
    2s20-50 & $\mathbf{0.013~ M_\odot \s^{-1}}$  & $2 \s$ & $20^\circ$         & $50^\circ$   &  $31^\circ$, $29^\circ$, $49^\circ$, $42^\circ$, $26^\circ$, $41^\circ$, $20^\circ$, $40^\circ$, $27^\circ$, $35^\circ$ & $160^\circ$, $153^\circ$, $132^\circ$, $155^\circ$, $133^\circ$, $147^\circ$, $155^\circ$, $144^\circ$, $145^\circ$, $143^\circ$ \\[0.4cm]
    1s20-50 & $\mathbf{0.026~ M_\odot \s^{-1}}$  & $1 \s$ & $20^\circ$         & $50^\circ$   &  same as above & same as above      \\[0.4cm]

    1s15-25 & $\mathbf{0.026~ M_\odot \s^{-1}}$  & $1 \s$ & $15^\circ$         & $25^\circ$    & $24^\circ$, $25^\circ$, $21^\circ$, $21^\circ$, $20^\circ$, $17^\circ$, $16^\circ$, $19^\circ$, $22^\circ$, $21^\circ$ & $160^\circ$,
$156^\circ$,
$160^\circ$,
$155^\circ$,
$165^\circ$,
$162^\circ$,
$162^\circ$,
$160^\circ$,
$159^\circ$,
$159^\circ$     \\
    \hline
    \end{tabular}
    \caption{The different runs {with jets} and their parameters. Angles are measured from the upper segment of the symmetry axis,
    $\dot M_j$ is the bipolar jets outflow rate (two jets together), and $t_j$ is the total operating time of the jets.
    $\theta_{\rm min}$ and $\theta_{\rm max}$ are the minimum and maximum angles of the allowed jets' angles
    (see Fig. \ref{fig:angles}), and $\theta_u$ and $\theta_d$ are the angles of the jet in the upper and lower halves of the grid, respectively.
    The launching velocity of all the jets is $100,000 \km \s^{-1}$.
     }
      \label{Tab:Table1}
\end{table}

\begin{figure}[h!]
\begin{center}
\includegraphics[width=0.65\textwidth]{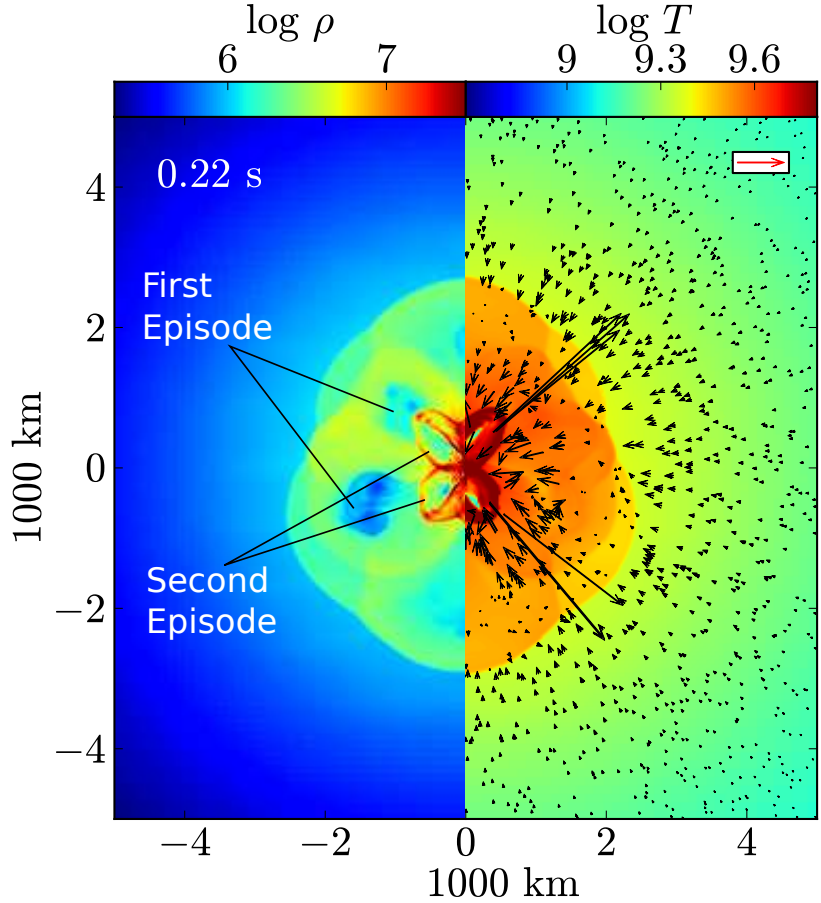}
\caption{The flow pattern during the second launching episode at time $t=0.22 \s$ of run 2s20-50. Marked are the bubbles formed by the jets.
Left: log density in $\g \cm^{-3}$. Right: log temperature in Kelvin. The arrows are the flow velocity with length proportional to the velocity
and a scale of $20,000 \km \s^{-1}$ given by the red arrow in the inset. Note that the bubbles formed by the first jets' episode are
flowing toward the center. }
\label{fig:global2sE}
\end{center}
\end{figure}

The results of run 2s20-50 at an early time, during the active phase of the second launching episode, are presented in Fig. \ref{fig:global2sE}.
The first jet on each side of the equatorial plane manages to penetrate through the infalling gas close to the center,
up to a distance of $\sim 1500 \km$ before it ceases to exist by our setting.
The jets of the first episode don't manage to reverse inflow, and by the time of the second episode shown in Fig. \ref{fig:global2sE}
the shocked jets' material can be seen falling inward.
For the first episode the angles of the upper and lower jets relative to the symmetry axis are
$\theta_u=31^\circ$, and $\theta_d=160^\circ$ ($20^\circ$ from the lower part of the symmetry axis in the figure), respectively.
Two new jets, one on each side of the equatorial plane, are launched $0.2 \s$ after the previous episode has started,
as in total we have ten episodes within 2 seconds.
The first small formed bubbles, one on each side of the equatorial plane (the horizontal plane in the figures),
and the newly forming bubbles from the second launching episode are marked and clearly seen in Fig. \ref{fig:global2sE}.

\begin{figure}[h!]
        \centering
        \begin{subfigure}{0.3\textwidth}
                \centering
                \includegraphics[width=\textwidth]{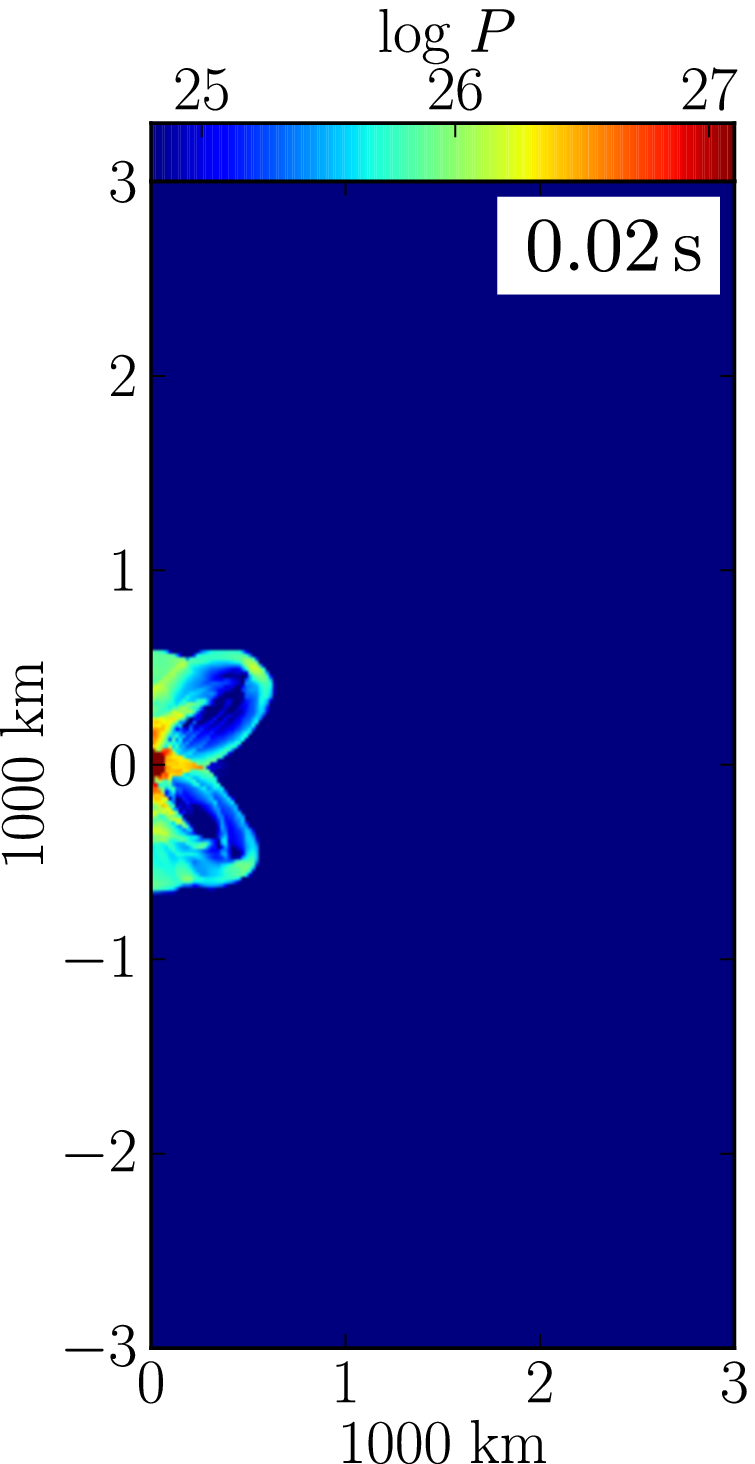}

        \end{subfigure}%
        ~ 
        \begin{subfigure}{0.3\textwidth}
                \centering
                \includegraphics[width=\textwidth]{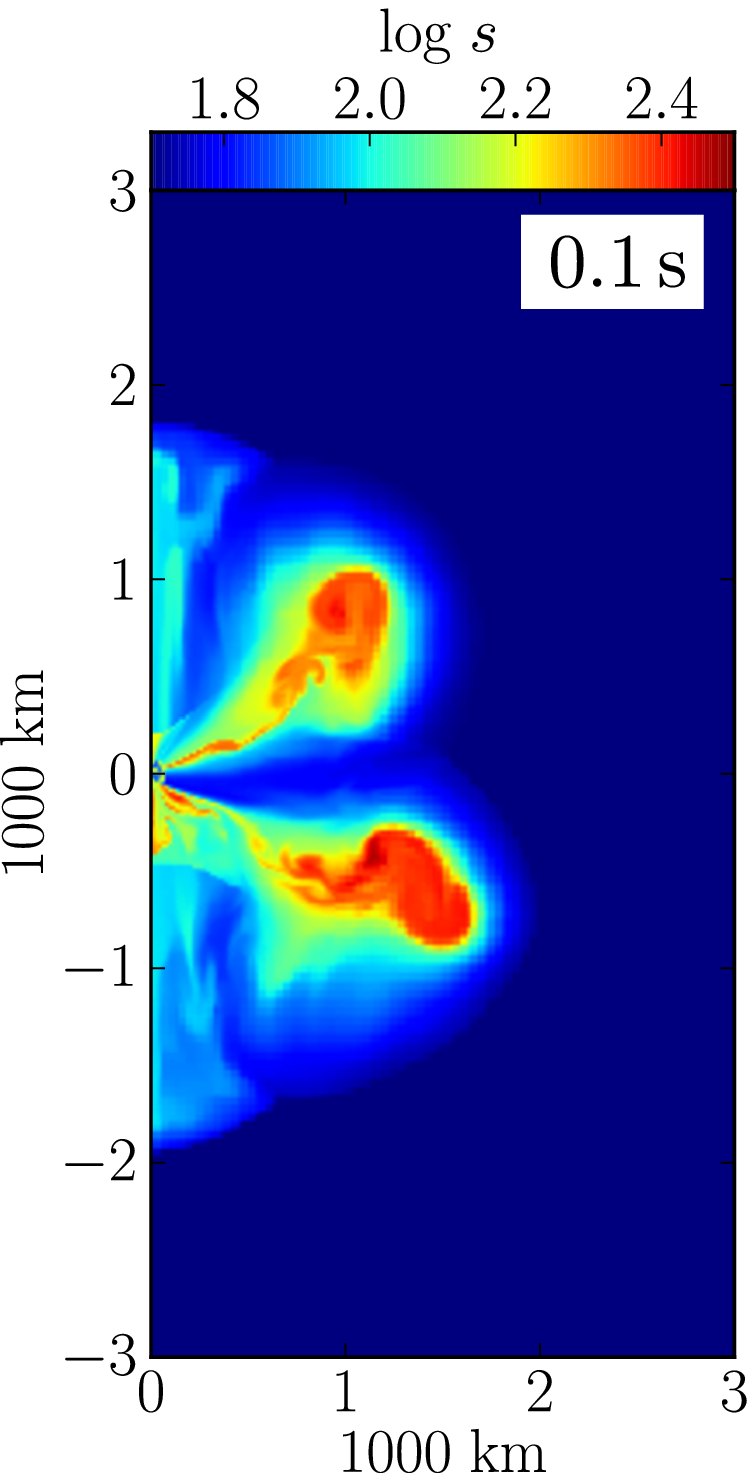}
        \end{subfigure}%
        ~ 
        \begin{subfigure}{0.3\textwidth}
                \centering
                \includegraphics[width=\textwidth]{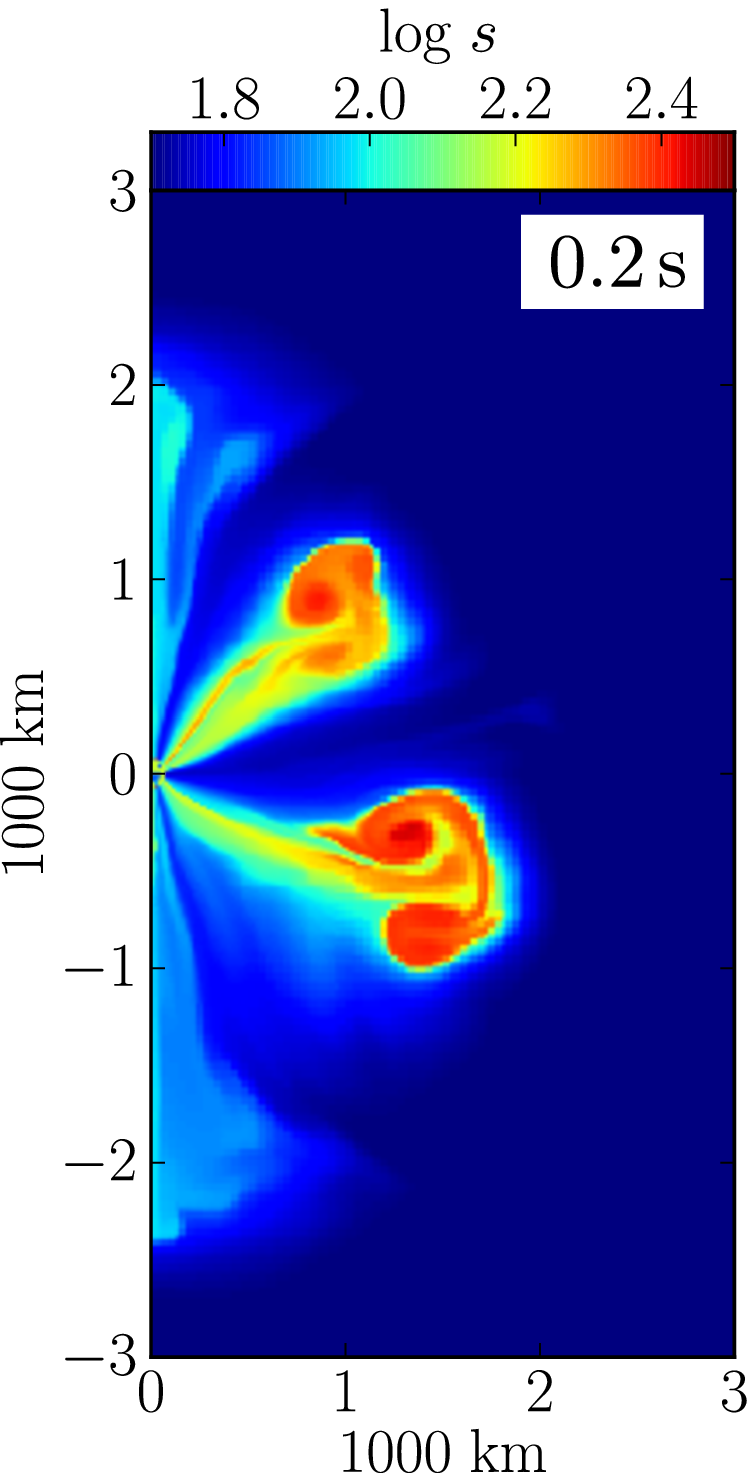}
        \end{subfigure}
        ~ 
 \caption{The flow of the first jets in run 2s20-50, emphasizing the limitation of the 2.5D grid. Left: log pressure in ${\rm dyn} \cm^{-2}$ .
  The pressure near the symmetry axis is clearly seen to be higher than near the equatorial plane, due to the limitation of the 2.5D grid (see text).
   Middle: log entropy {in units of $k_{\rm B}$ per baryon}  of the same run at a later time of $t=0.1$. The jets now have different directions than their initial ones, and they
   are bent toward the equatorial plane. Right: log entropy in units { of $\rm k_B$ per baryon} at a still later time of $t = 0.2 \s$, just before the second episode is launched.}
\label{fig:bend}
\end{figure}

Another characteristic of the first two jets, one on each side of the equatorial plane, is that they are bent away from the symmetry axis.
This is clearly seen in Fig. \ref{fig:bend}. This effect is mainly a numerical effect because of the 2.5D nature of the grid.
As we actually launch a conical shell, the shocked jets and ambient gas form very dense region closer to the axis.
On the other side, toward the equatorial plane, the shocked gas has large volume to expand to. A large pressure is built near the axis
as can be seen in the left panel of Fig. \ref{fig:bend}. This pressure pushes the jets toward the equatorial plane, and
the jets expand almost parallel to the equatorial plane, as seen in the middle panel of Fig. \ref{fig:bend}.
The right panel of Fig. \ref{fig:bend} is at a time after the first jets have been turned off, and the bubbles are pushed
back toward the center by the inflowing ambient gas.
In case of several launching episodes near a defined axis, the numerical effect of high pressure near this axis becomes
real even in 3D simulations. Full 3D simulations are required to study the flow more accurately.

\begin{figure}[h!]
        \centering
        \begin{subfigure}{0.5\textwidth}
                \centering
                \includegraphics[width=\textwidth]{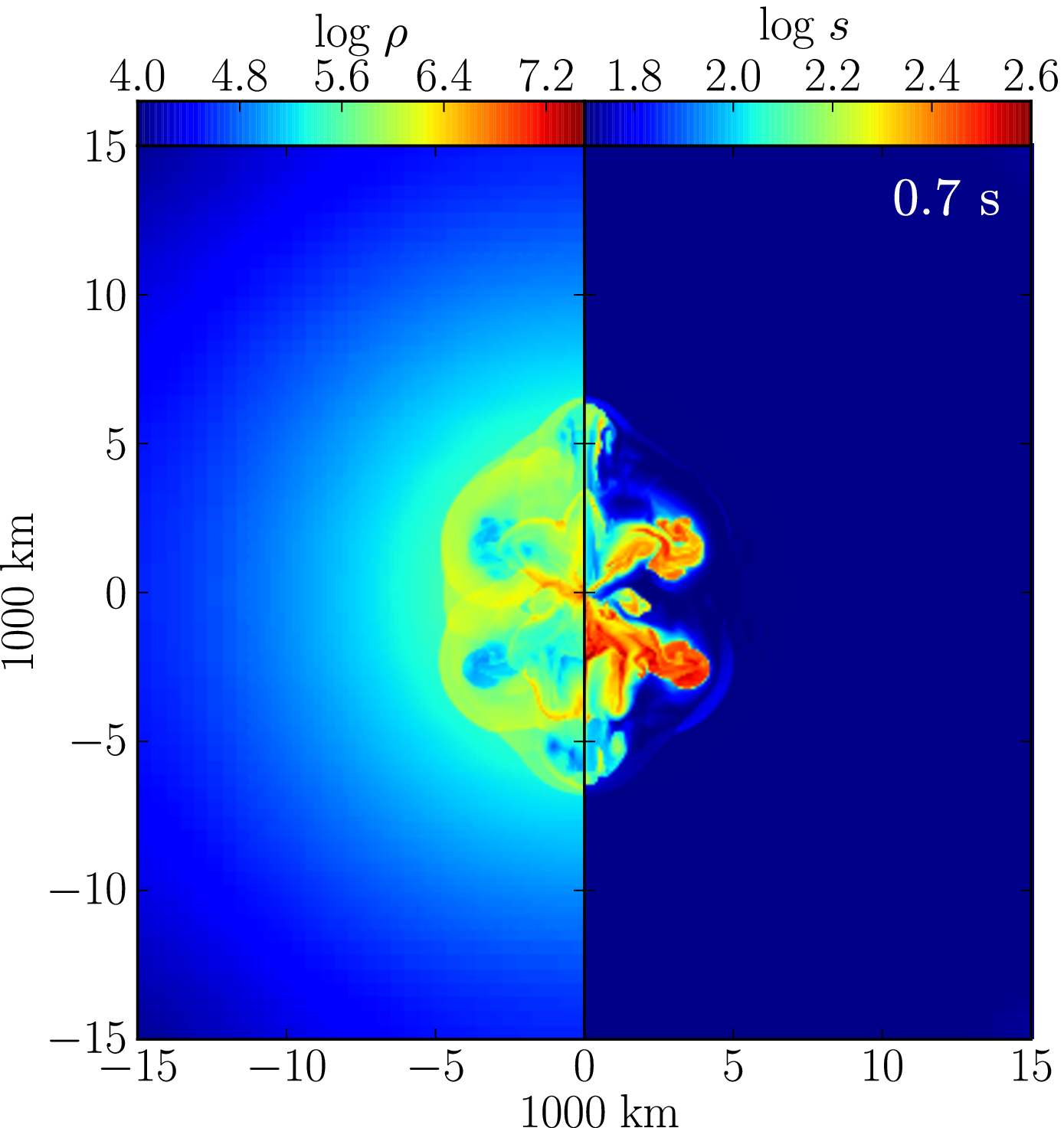}
                \label{fig:2s_t070}
        \end{subfigure}%
        ~ 
        \begin{subfigure}{0.5\textwidth}
                \centering
                \includegraphics[width=\textwidth]{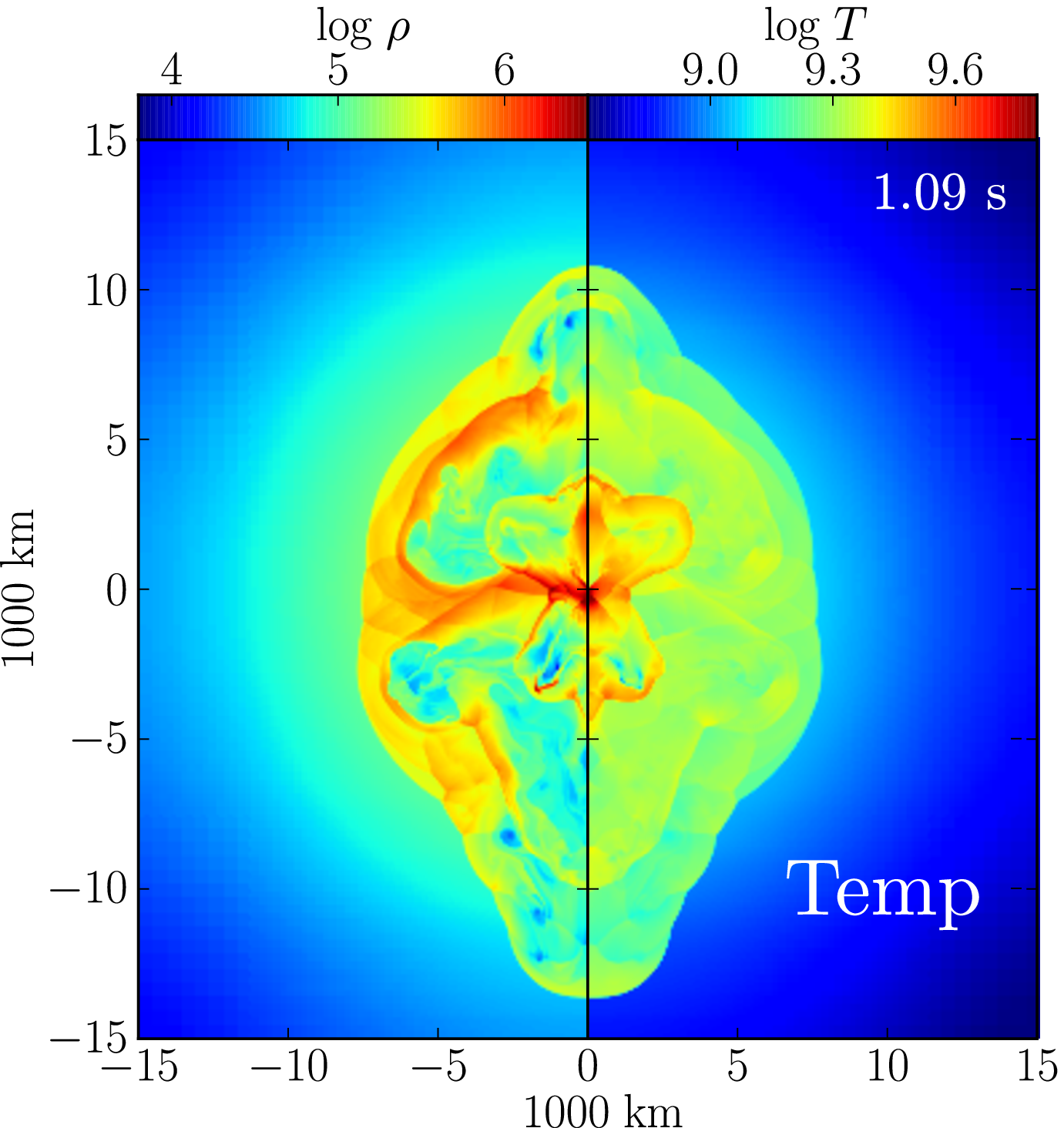}
                \label{fig:2s_t109}
        \end{subfigure}
        ~ 
        \begin{subfigure}{0.5\textwidth}
                \centering
                \includegraphics[width=\textwidth]{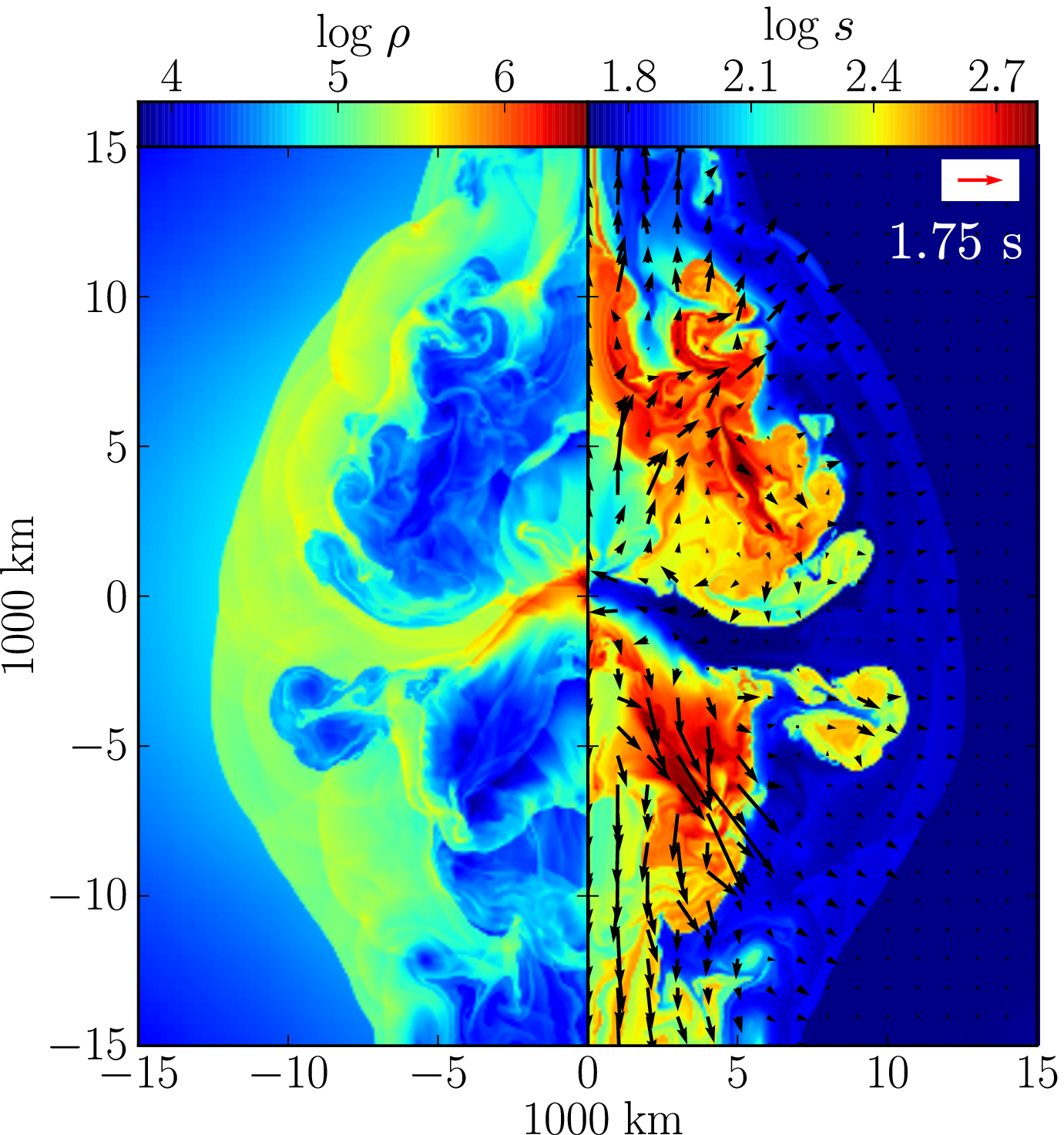}
                \label{fig:2s_t109}
        \end{subfigure}
        \caption{Results of run 2s20-50 at three times. The left half of all panels shows the density, with a color coding in logarithmic scale and units of $\g \cm^{-3}$.
        The right side at times $t=0.7\s$, $t = 1.09 \s$, and $t = 1.75 \s$ show the { log entropy in units of $\rm k_B$ per baryon}, temperature in log scale in units of $\K$, and the flow velocity.
        The entropy map emphasizes the shocked jets' material. The arrows are the flow velocity with length proportional to the velocity
and a scale of $20,000 \km \s^{-1}$ given by the red arrow in the inset.}
        \label{fig:global2s}
\end{figure}

The evolution at later times is presented in Fig. \ref{fig:global2s}.
The left half-plane of all three panels presents the density map, while the right half-plane shows the entropy ($t=0.7\s$), the temperature
($t = 10.9 \s$), and the velocity map ($t = 1.75 \s$).
Each new jet propagates somewhat faster as it plows its way through the lower density regions left behind by the previous jets.
Each jet further shocks the gas and increases the volume of the low-density high-temperature region.
The series of jets leads to the formation of small bubbles that eventually merge to form one bubble on each half of the
equatorial plane. Later these two bubbles are likely to merge at some distance from the center to build one large bubble that pushes
the core material and forms a more spherical outflow.
In the 2.5D simulation this process takes longer than our simulation time, because each region in the plane is actually a ring.
In any case, these two bubbles accelerate the core material outward; this is the onset of the explosion.
During the simulations the temperature in the bubbles remains in the range of $2-5 \times 10^9 \K$.
This temperature has implications to the r-process that might occur in the bubbles \citep{Papish2012b}.
A prominent feature is that mass accretion proceeds from regions near the equatorial plane. This is analyzed in section \ref{sec:acc}.

\begin{figure}[h!]
        \centering
        \begin{subfigure}{0.5\textwidth}
                \centering
                \includegraphics[width=\textwidth]{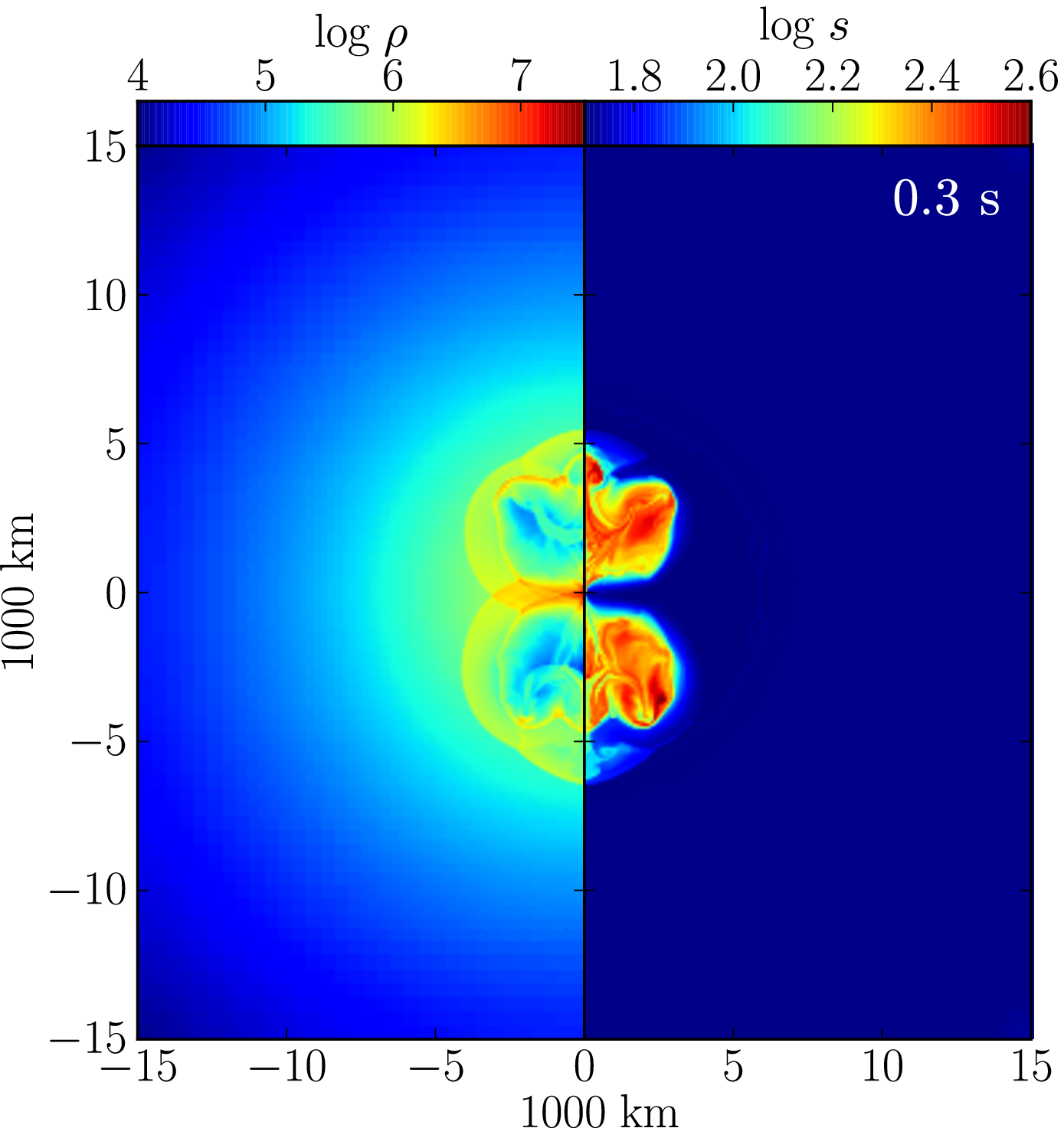}
                \label{fig:1s_t070}
        \end{subfigure}%
        ~ 
        \begin{subfigure}{0.5\textwidth}
                \centering
                \includegraphics[width=\textwidth]{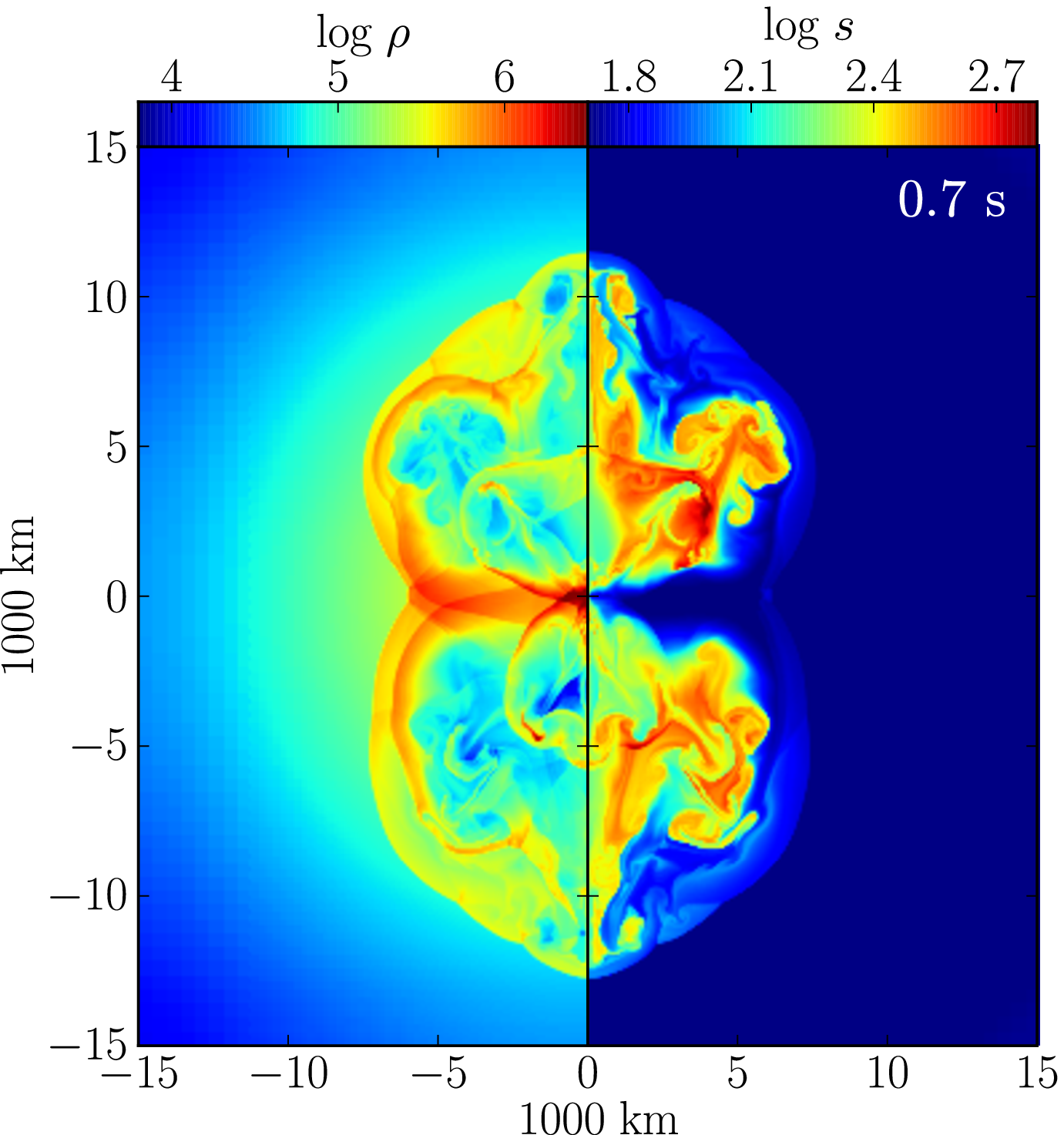}
                \label{fig:1s_t070}
        \end{subfigure}%
        \\
        ~ 
        \begin{subfigure}{0.5\textwidth}
                \centering
                \includegraphics[width=\textwidth]{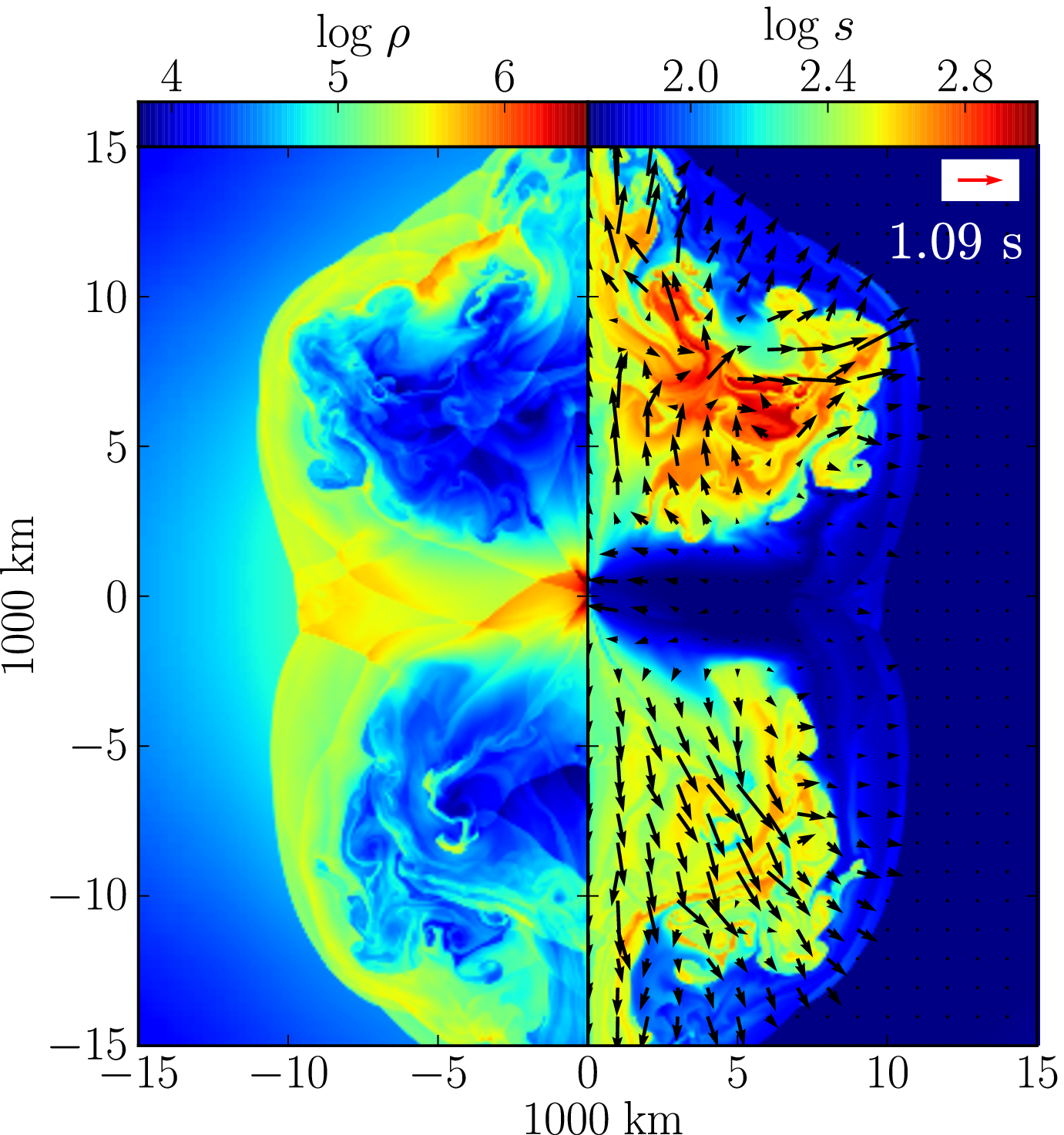}
                \label{fig:1s_t109}
        \end{subfigure}
        ~ 
        \caption{Results of run 1s20-50 at three times. The jets are for a total duration of $1 \s$, rather than $2 \s$ as in Fig. \ref{fig:global2s}. At $t=1.09 \s$ material near the equatorial plane still flows in and feeds the center, but further out the bubbles are pushing out material from the equatorial plane.
        Left and right sides of each panel show the log density { in $\g \cm^{-3}$ and log entropy in units  of $\rm k_B$ per baryon}. The arrows are the flow velocity with length proportional to the velocity
and a scale of $20,000 \km \s^{-1}$ given by the red arrow in the inset.}
         \label{fig:global1s}
\end{figure}

The structure of the final outflow depends on the properties of the jets, as dictated in our model by several parameters.
In the present study we vary only the duration of the total activity time and the jittering boundaries of the jets.
In Fig. \ref{fig:global1s} we present the results of run 1s20-50 that has the same jittering boundaries and the same total injected
mass and energy.
The number of episodes stays at ten, but each episode, the time between episodes, and the total activity time are half that of
run 2s20-50.
The main differences between the two runs are as follows.
(1) The more powerful and shorter activity of the jets in run 1s20-50 leads to a more spherical outflow at early times.
(2) The inflow region near the equatorial plane is wider in the shorter-episodes 1s20-50 run.
Other than these the two runs will lead to similar explosions.

\begin{figure}[h!]
        \centering
        \begin{subfigure}{0.5\textwidth}
                \centering
                \includegraphics[width=\textwidth]{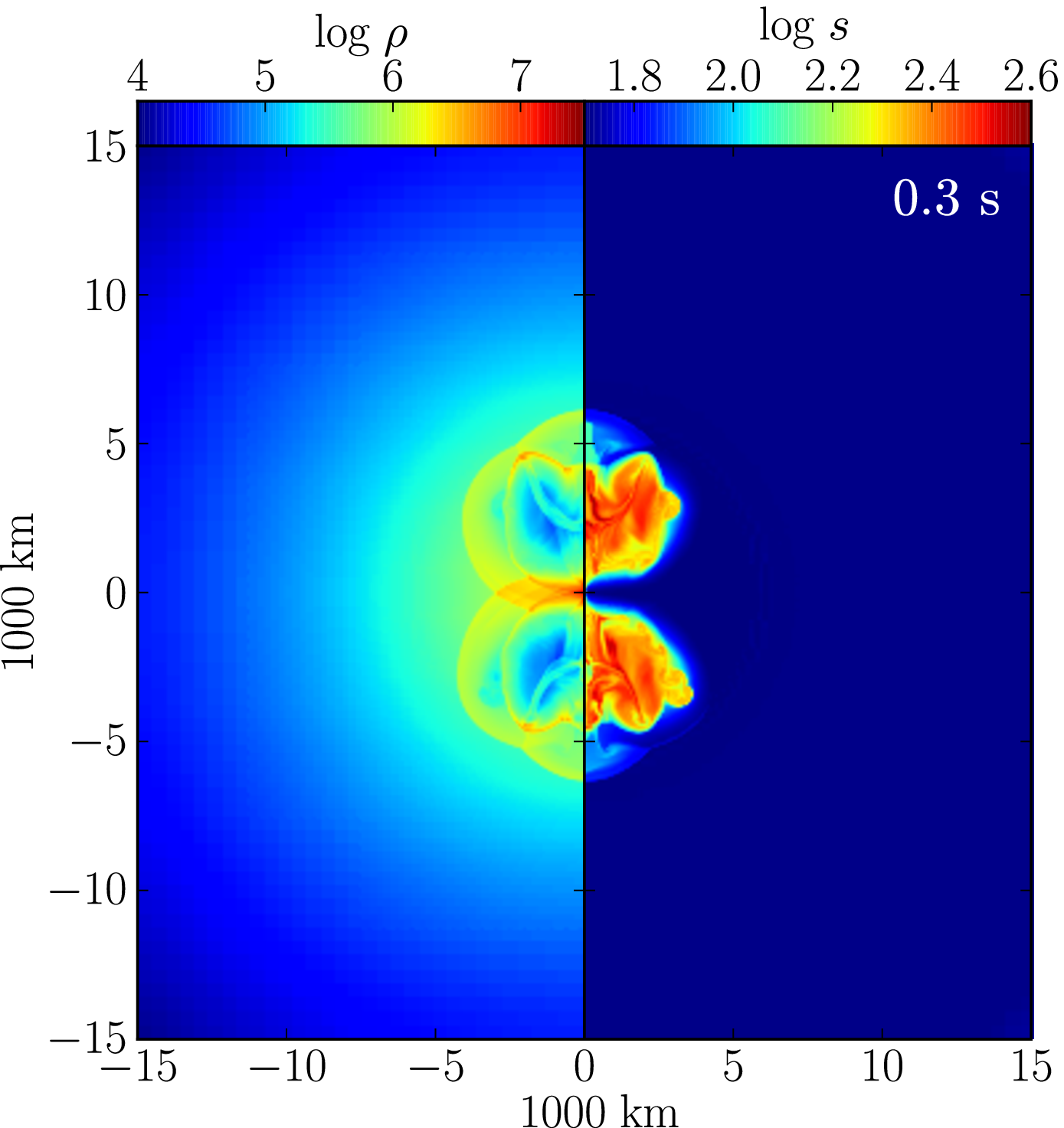}
                \label{fig:1s15-25_t030}
        \end{subfigure}%
        ~ 
        \begin{subfigure}{0.5\textwidth}
                \centering
                \includegraphics[width=\textwidth]{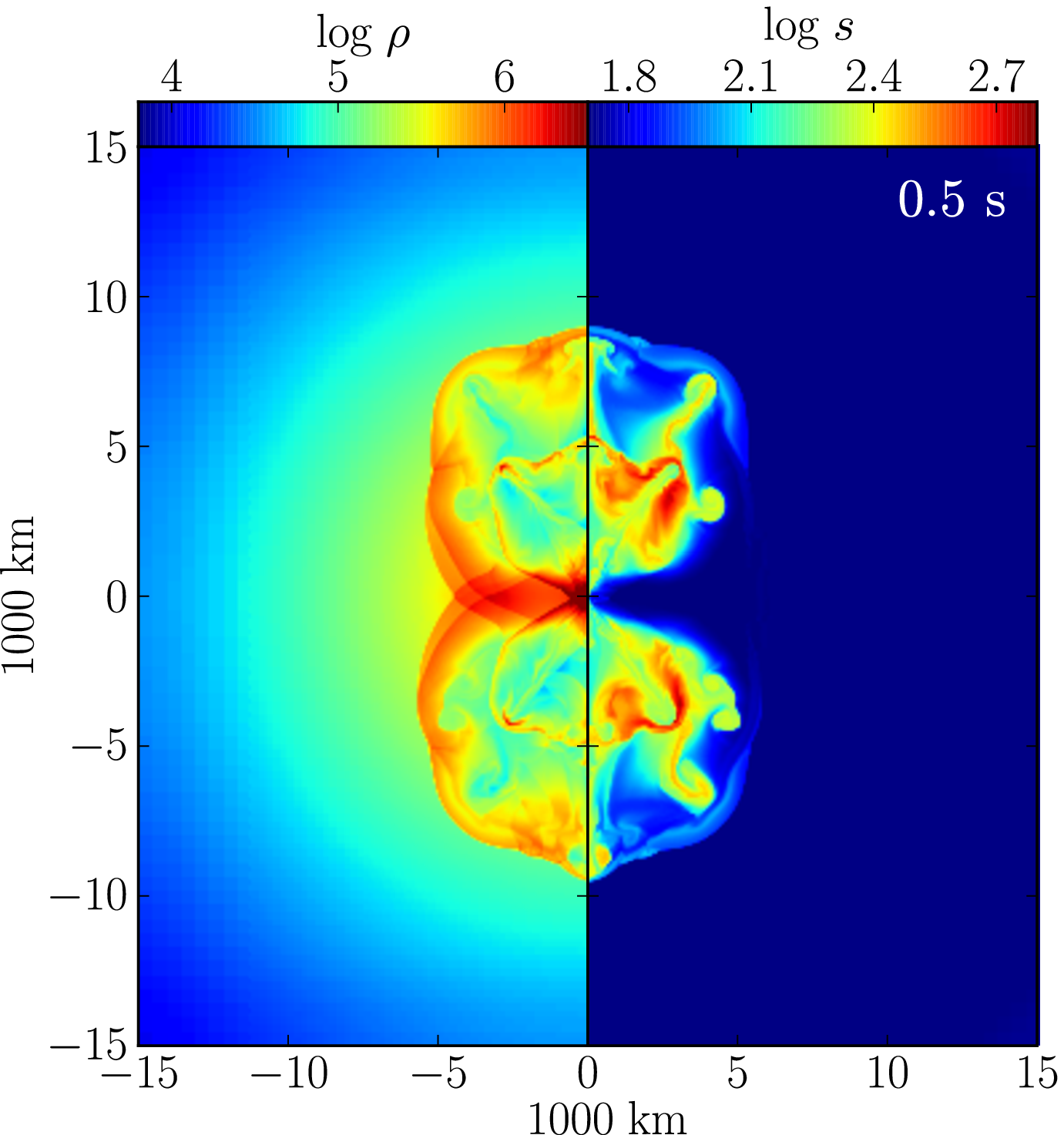}
                \label{fig:1s15-25_t050}
        \end{subfigure}%
        ~ 
          \\
        \begin{subfigure}{0.5\textwidth}
                \centering
                \includegraphics[width=\textwidth]{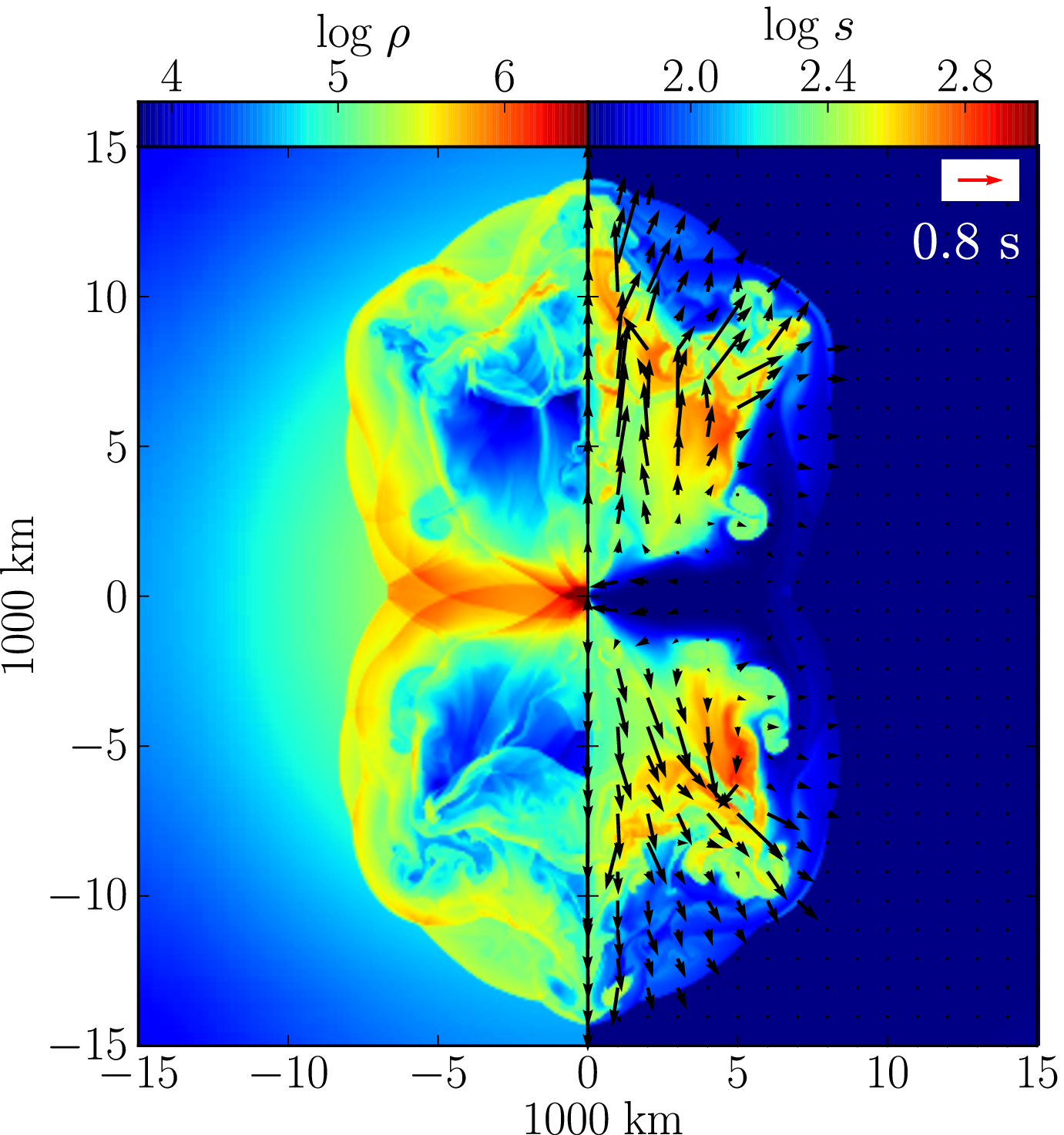}
                \label{fig:1s15-25_t080}
        \end{subfigure}
        ~ 
        \caption{Like Fig. \ref{fig:global1s} but for run 1s15-25. }
         \label{fig:global1s15-25}
\end{figure}

In run 1s15-25 we change the boundaries of the jittering to be narrower and closer to the symmetry axis,
$\theta_{\rm min}=15^\circ$ and $\theta_{\rm max}=25^\circ$, but all other parameters are as in run  1s20-50.
The flow properties are presented in Fig. \ref{fig:global1s15-25}.
The main result is that even that the jets are launched close to the symmetry axis the two large bubbles manage to accelerate
material close to the equatorial plane.
Part of the effect is due to the bending of the jets away from the symmetry axis as discussed above.
{{{{ In a case of small angular momentum in the pre-collapse core one can define a symmetry axis along the angular momentum axis. Jets are likely to be launched close to this axis, and the built-up of pressure near
the symmetry axis is real even in a 3D flow.}}}}
To find the exact behavior of the jets we are planning 3D numerical simulations.
Even if quantitatively not accurate (because of the 2.5D nature of the grid), the result of this run is significant,
as it shows that there is no need for jittering to be over the entire sphere in order to lead to an efficient explosion.

\begin{figure}[h!]
        \begin{subfigure}{0.5\textwidth}
                \includegraphics[width=\textwidth]{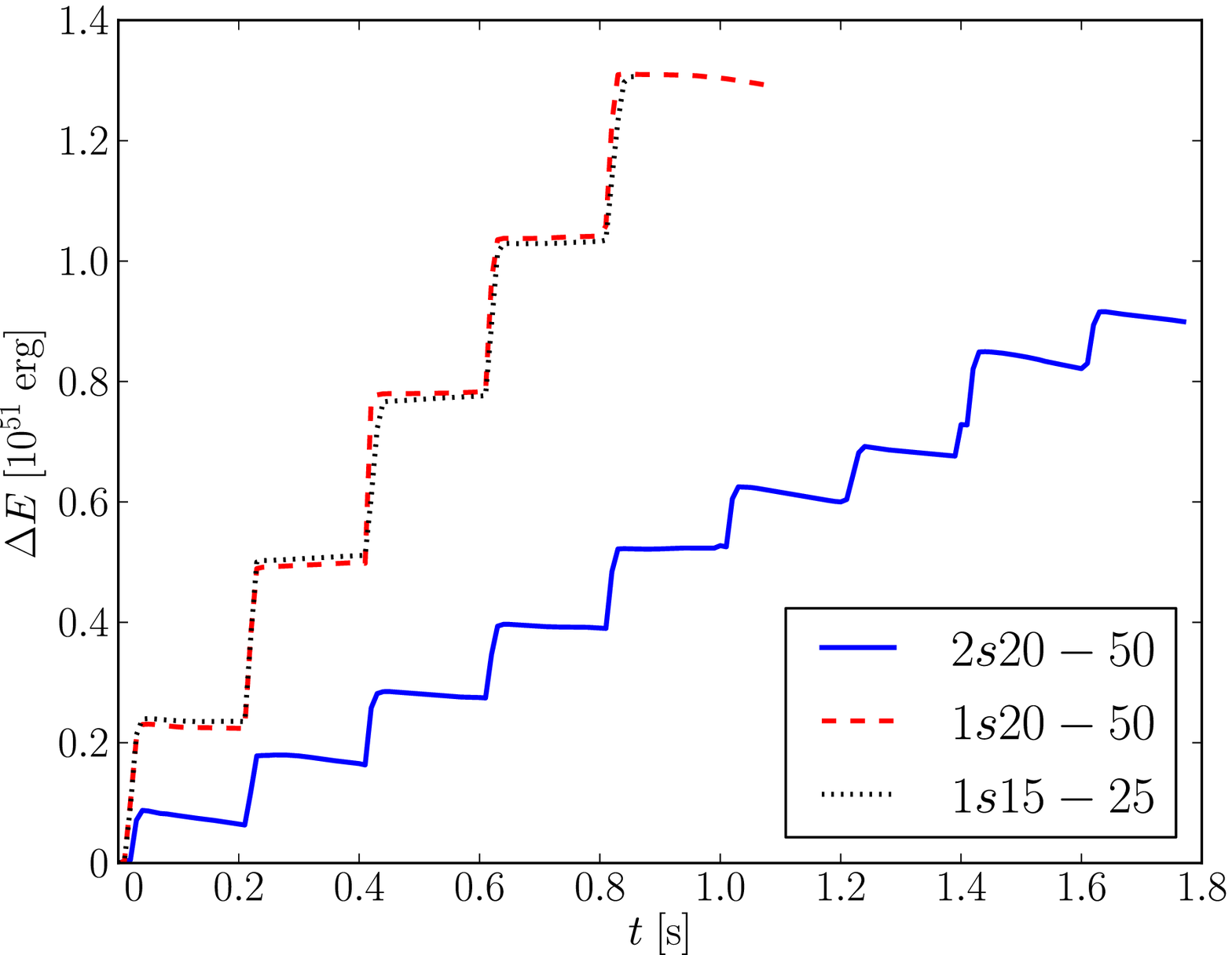}
        \end{subfigure}
                       \begin{subfigure}{0.5\textwidth}
                \includegraphics[width=\textwidth]{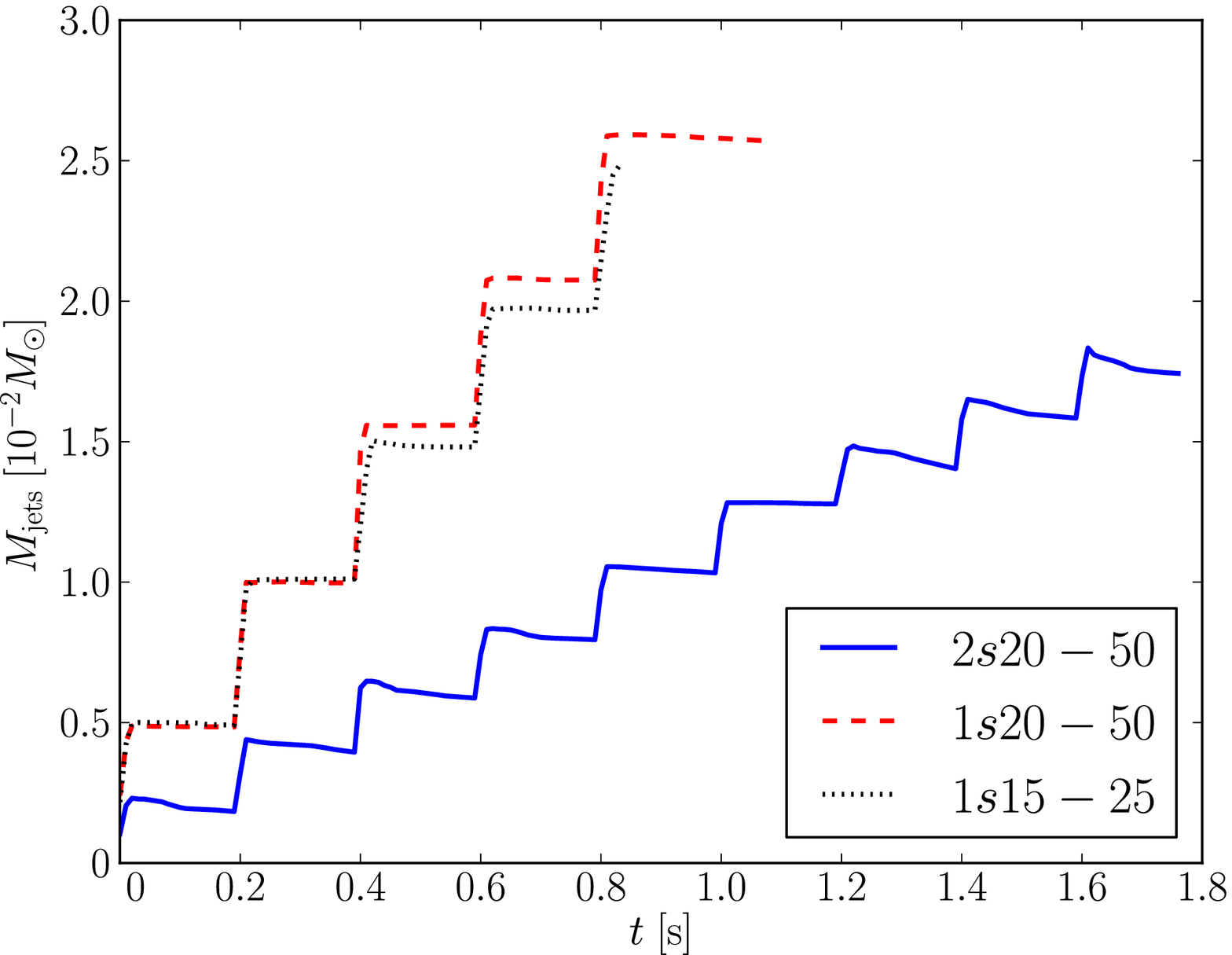}
        \end{subfigure}%
        ~ 
        \caption{{Left: The evolution of the total energy $\Delta E$ (kinetic+thermal+gravitational) relative to the total energy at the beginning of the simulation for the three different models. Right: The evolution of the total jets material for the three models.  }}
        \label{fig:global5}
\end{figure}

{{{ {Finally we present the total mass and energy in the outflows for the three models in Fig. \ref{fig:global5}. The total energy and mass are calculated up to the time when the outflows get out of the grid.
For model $2s20-50$ with weaker jets it is shown that part of the outflow is accreted back to the center while for the stronger jets in models $1s20-50$, and $1s15-25$ there is almost no backflow to the origin.}
}}}
\subsection{Small Scale Features}
\label{subsec:local}

\begin{figure}[h!]
\begin{center}
\includegraphics[width=0.5\textwidth]{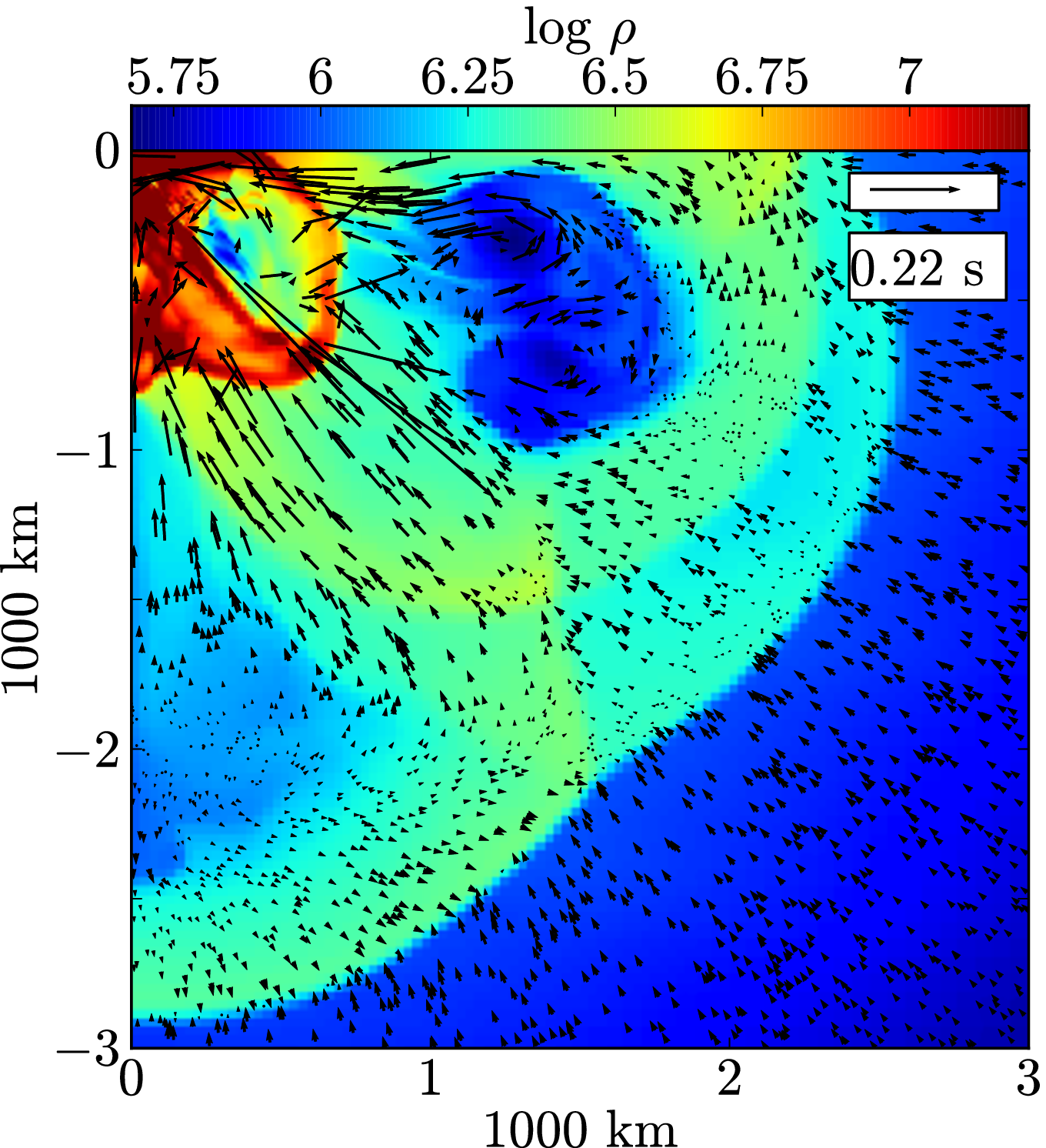}
\caption{Zoom-in of Fig. \ref{fig:global2sE} at time $t = 0.22 \s$. The arrows are the flow velocity with length proportional to the velocity
and a scale of $20,000 \km \s^{-1}$ given by the arrow in the inset.
Two opposite vortices are present in the upper bubble.}
\label{fig:zoom}
\end{center}
\end{figure}

The interaction of jets with an ambient medium that leads to bubble inflation has some generic features.
In particular, vortices are formed inside and outside the bubbles \citep{Soker2013}.
These vortices might mix the ambient medium with the shocked jets' material, and the shocked jets' material itself within the
bubbles.
In Fig. \ref{fig:zoom} we show the flow structure of the isolated bubble seen in the lower left part
of Fig. \ref{fig:global2sE}. The vortices inside the bubble are clearly seen. The first jet creates two opposite vortices while material around the bubble is keeps falling in.
{{{ {The small-scale features of the flow, like vortices, are more complicated in 3D. This will not influence much the inflation of
the bubble, as the volume of the bubbles is determined mainly by energy conservation. However, the changes in local thermodynamics properties are expected to affect the outcome of the r-process. }}}}
This mixing within hot bubbles has implications for nuclear synthesis within the bubbles,
particularly possible r-process elements \citep{Papish2012a}.
This topic is beyond the goals of the present study.

\begin{figure}[h!]
        \begin{subfigure}{0.5\textwidth}
        \hspace{-1.5 cm}
                \includegraphics[width=\textwidth]{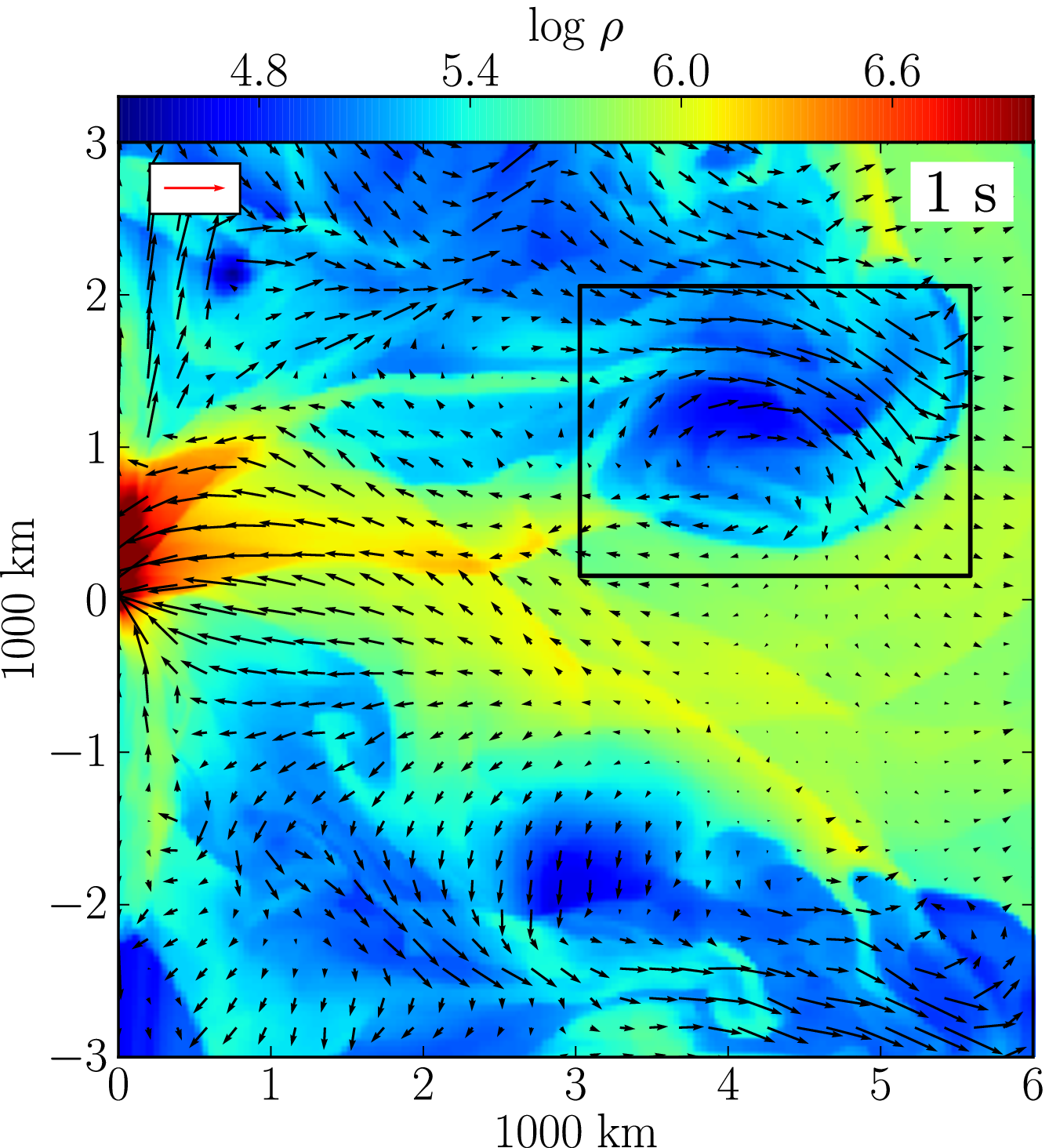}
        \end{subfigure}
                \begin{subfigure}{0.62\textwidth}
                \hspace{-1. cm}
                \includegraphics[width=\textwidth]{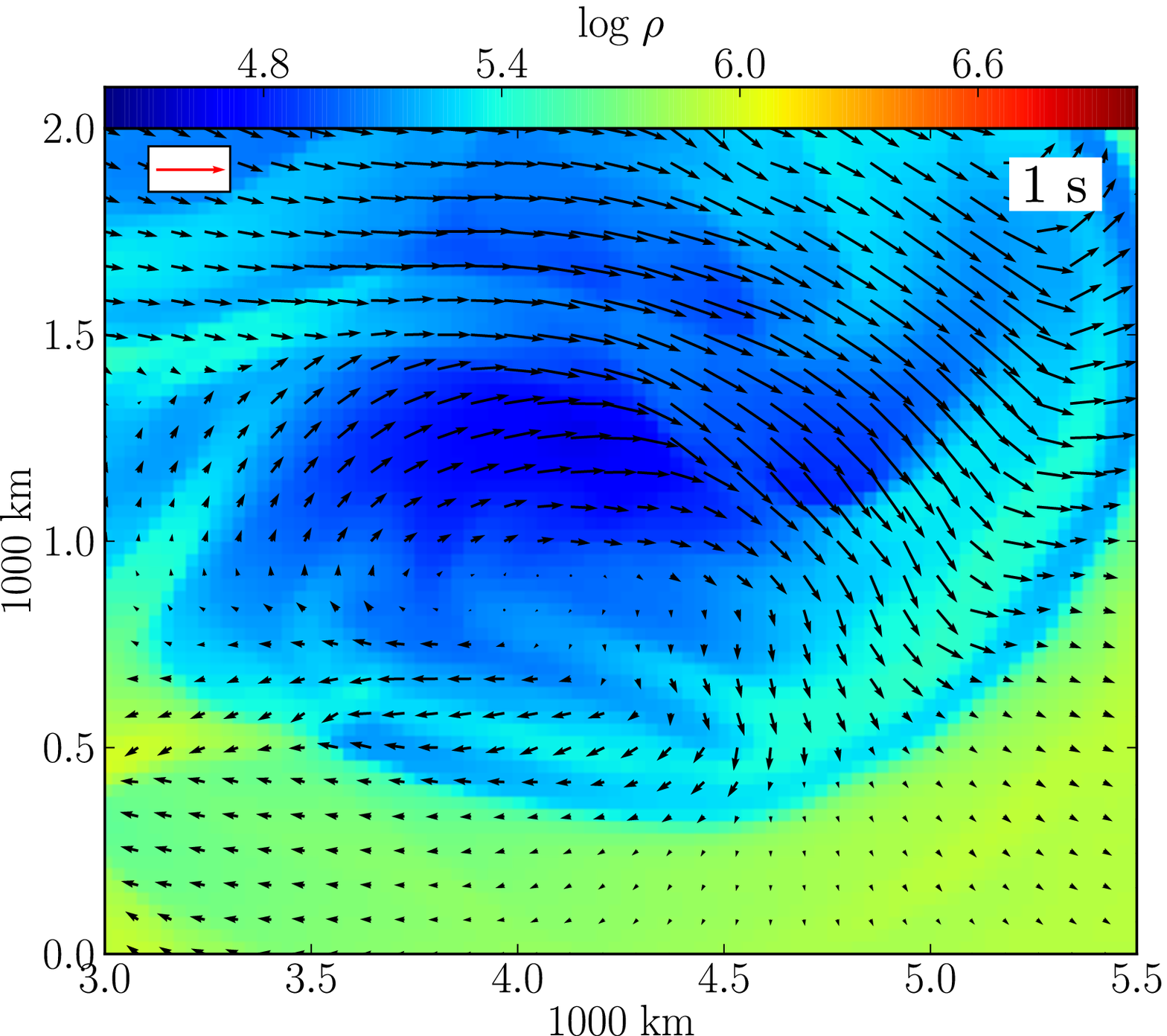}
        \end{subfigure}%
        ~ 
        \caption{Shown are zoom-in of run 2s20-50 (Fig. \ref{fig:global2s} at time $t = 1.0 \s$. The  left panel shows the log density in cgs units. The arrows are the flow velocity with length proportional to the velocity and a scale of $20,000 \km \s^{-1}$ given by the arrow in the inset. Right: farther zoom-in of the black box marked on the left panel. The vortex is clearly seen.}
        \label{fig:t2_local}
\end{figure}

In Fig. \ref{fig:t2_local} we show the flow structure of run 2s20-50 at a later time when the flow is more complicated and
more vortices are present.
Vortices are formed where the shear is large. In the cases presented in Fig. \ref{fig:t2_local} the large vortices are formed between the inflow
close to the equatorial plane and the outflow of jets and bubbles.
The right panel in the figure emphasizes the structure of one vortex.

\subsection{Comparison with Non-Jittering Jets}
{{{
{
To study the importance of the jets' jittering to the flow structure we compare our results to two  cases with non-jittering bipolar jets: (a) constant jets injected in the direction of the symmetry axis and (b) constant jets injected at $50^\circ$ to the symmetry axis.}

{For case (a)  we take the jets to have the same opening angle and total energy as in models $1s20-50$ and  $1s15-25$. The results are shown in Fig. \ref{fig:axial}. The jets are able to penetrate the surrounding gas without bending while material from the equatorial plane is accreted to the center. The jets are not able to stop the accretion during their activity. We conclude that no feedback mechanism is present and the accretion will continue until most of the material from the equatorial plane is accreted. 
}

{In non-jittering case (b) we inject the jets until their total energy is equal to that in model $1s20-50$ ($\sim 0.12 \s$). We run the simulation farther up to the time when  the jets get out of the simulation grid. The results at time $t=0.47 \s$ are shown in Fig. \ref{fig:non-jittering}. The jets are able to expel material from the equatorial plan but are much less efficient in expelling material from directions close to the symmetry axis resulting in a higher accretion rate. As a conical outflow can be though as a rapidly precessing jet, this case is a limiting case of the jittering simulations.
\begin{figure}[h!]
        \centering
                \begin{subfigure}{0.5\textwidth}
                \includegraphics[width=\textwidth]{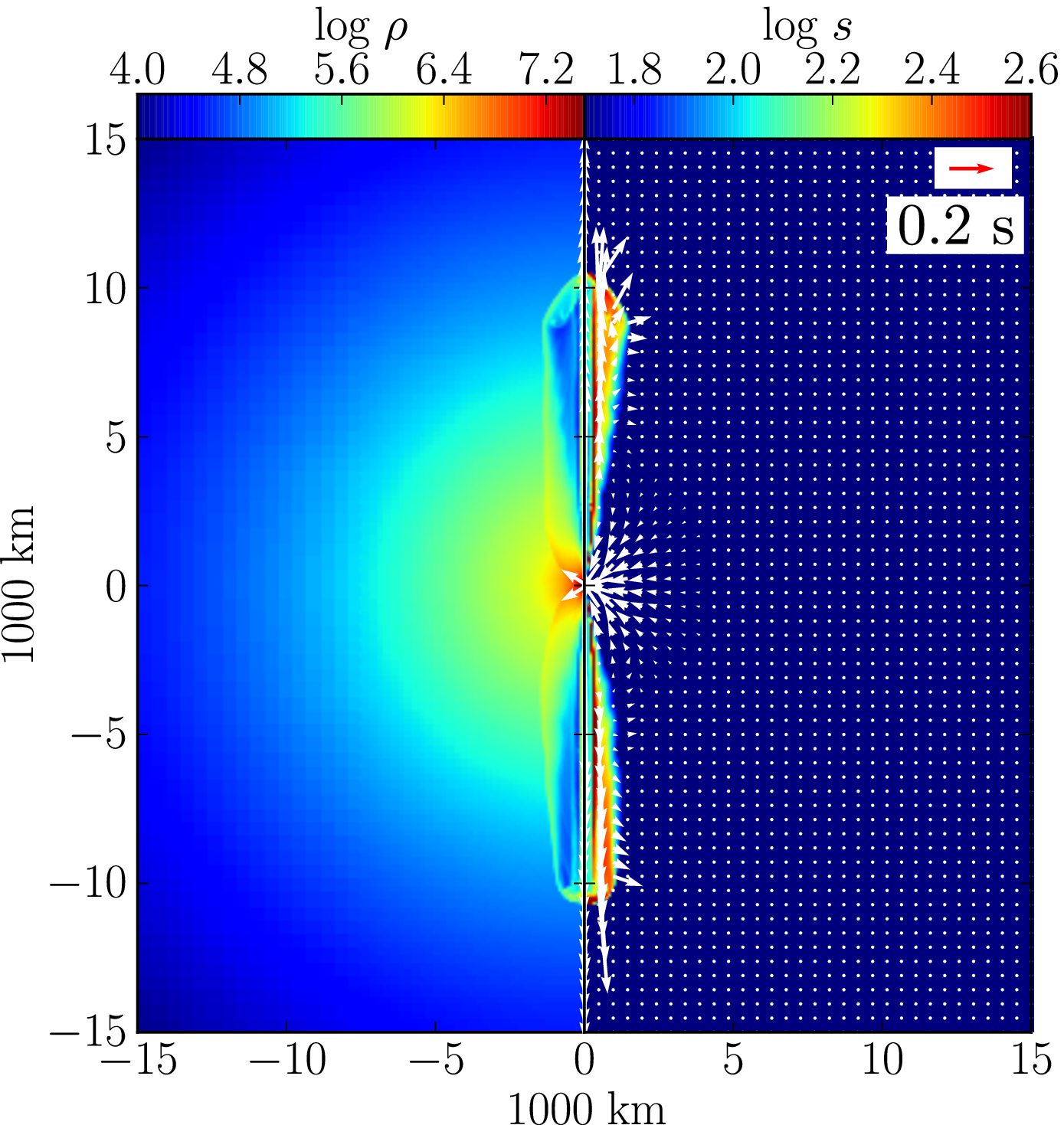}
        \end{subfigure}
        \caption{{Jets injected in the direction of the symmetry axis without jittering. The left panel shows the log density in $\g \cm^{-3}$; the right panel shows the log entropy in units of $k_{\rm B}$ per baryon. The arrows are the flow velocity with length proportional to the velocity and a scale of $20,000 \km \s^{-1}$ given by the arrow in the inset.}}
        \label{fig:axial}
\end{figure}
\begin{figure}    
        \begin{subfigure}{0.52\textwidth}
                \includegraphics[width=\textwidth]{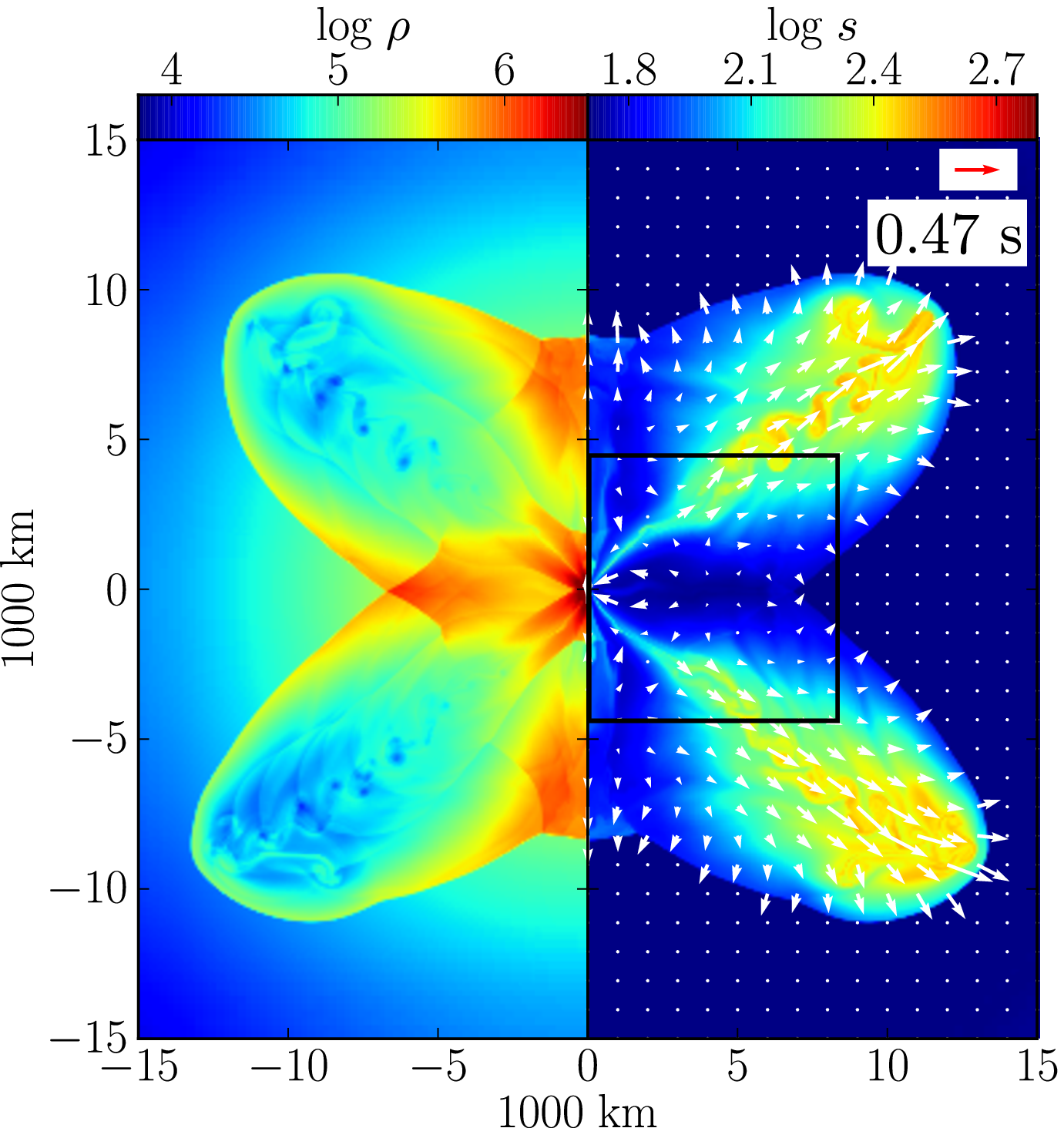}
        \end{subfigure}%
                \begin{subfigure}{0.5\textwidth}
                \includegraphics[width=\textwidth]{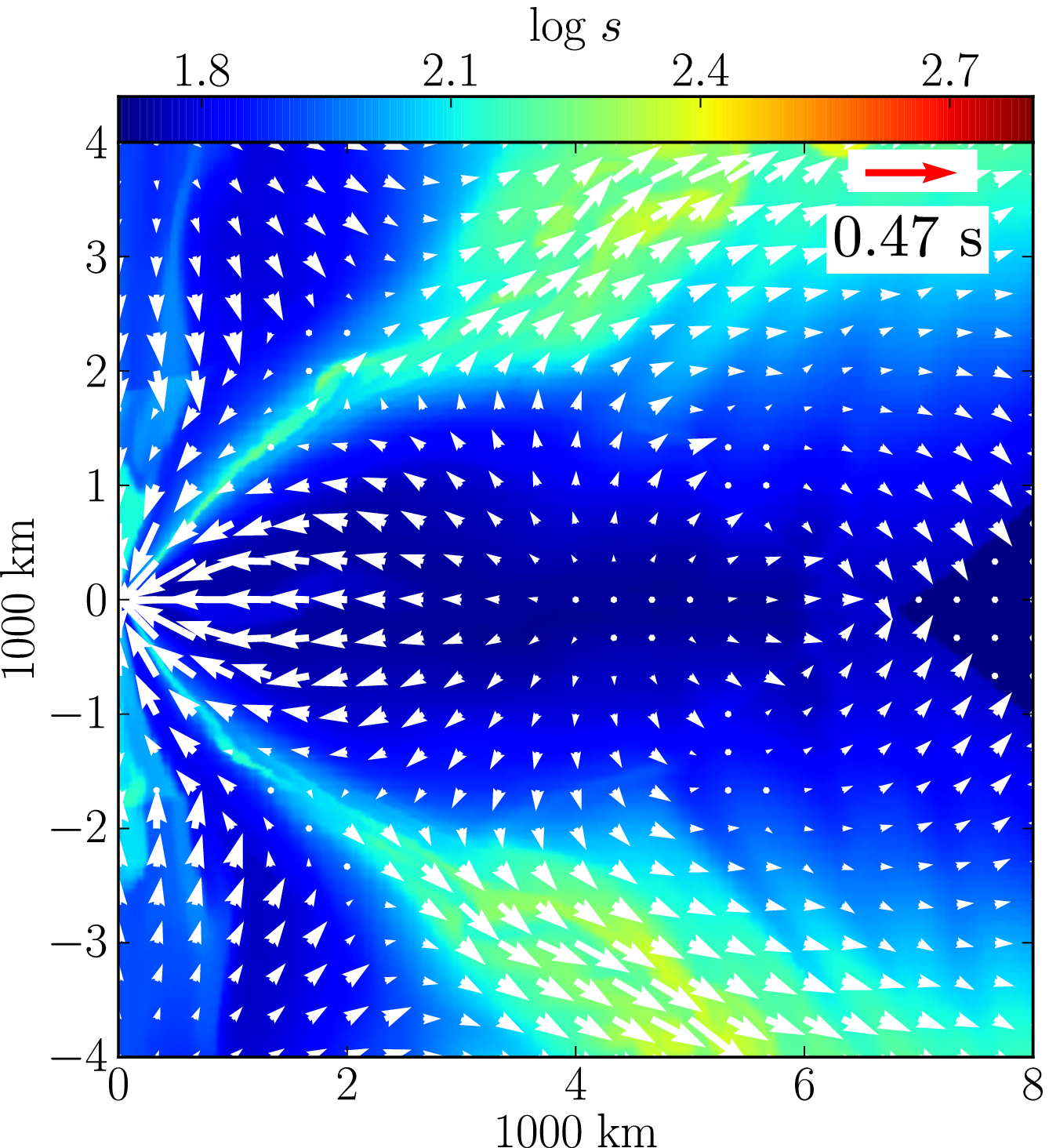}
        \end{subfigure}%
        ~ 
        \caption{{Jets injected in constant direction of $50^\circ$ relative to the symmetry axes without jittering. The left side of the left panel shows the log density in $\g \cm^{-3}$; the right side shows the log entropy in units of $k_{\rm B}$ per baryon. The arrows are the flow velocity with length proportional to the velocity and a scale of $20,000 \km \s^{-1}$ given by the arrow in the inset. The right panel shows the zoom-in of the black box marked on the left figure. The inflow along the symmetry axis is clearly seen.}}
        \label{fig:non-jittering}
\end{figure}
}}}}
\section{FEEDBACK EXPLOSION MECHANISM}
\label{sec:acc}

In the jittering-jet explosion model the feeding of the accretion disk that launches the jets
is part of a feedback cycle.
As long as the jets don't manage to explode the entire core the accretion proceeds.
When ejection is less efficient, e.g., when the jets are not jittering, accretion continues for a longer time.
The total energy of the jets must be larger than the binding energy of the core to halt accretion.
This feedback cycle accounts for the F.O.E. (fifty one ergs) typical CCSN explosion energies.
The feedback mechanism of the jittering-jet explosion model also implies that
\emph{less efficient jets-core interactions result in more energetic explosions.}
For example, a binary companion in a common envelope with the progenitor can eject the hydrogen and helium layers of the star
and spin-up the core. The rapidly rotating collapsing core, in the jittering-jet model, launches well-collimated jets
that don't expel the gas from near the equatorial plane. A more massive NS will be formed, with much more energy
carried by the jets. In extreme cases a BH will be formed.

\begin{figure}[h]
\begin{center}
\includegraphics[width=0.5\textwidth]{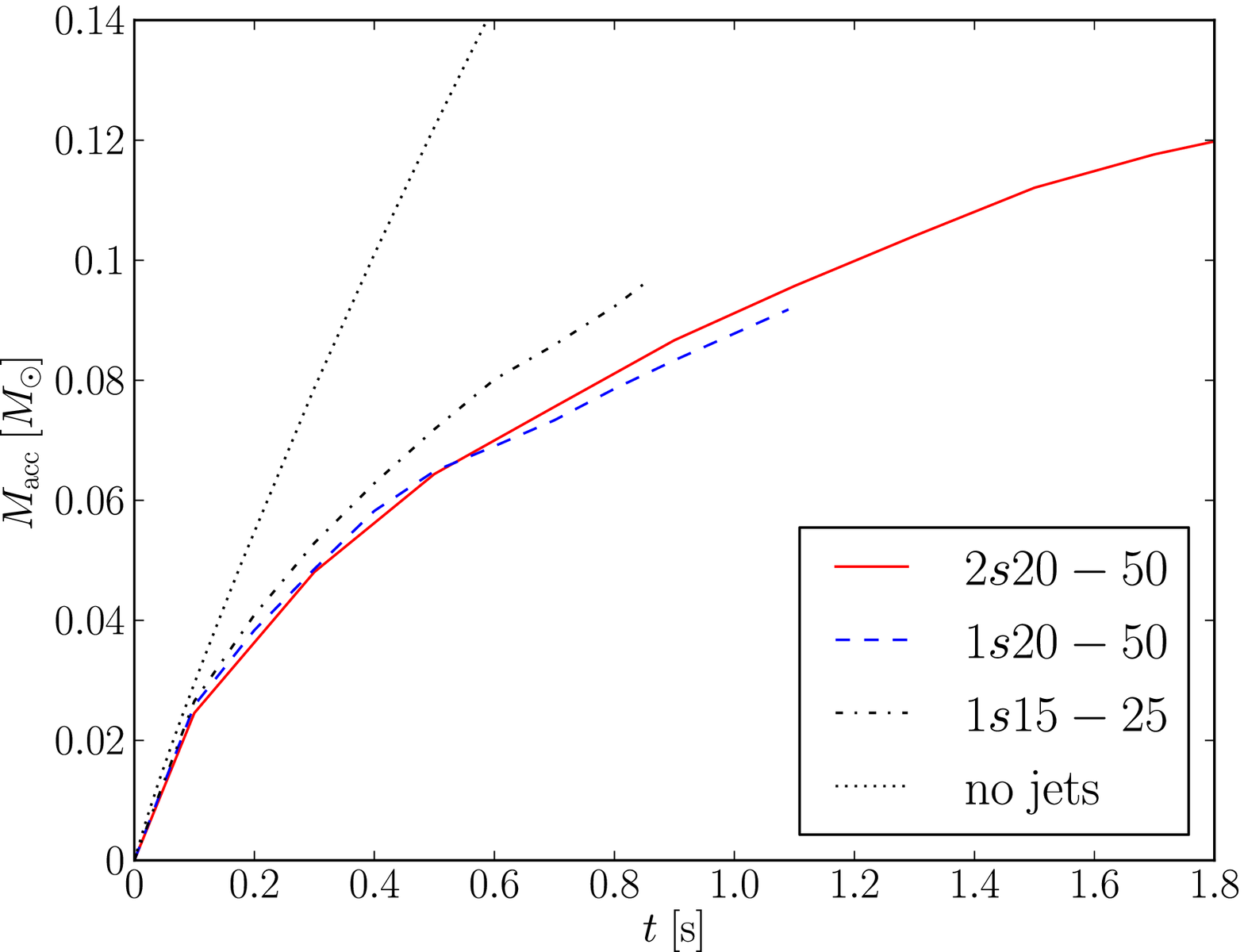}
\caption{The total accreted mass as function of time for the three cases {with jets} studied here {and for a case without jets for comparison}.
 Accreted mass is the mass that enters a spherical surface of radius $75 \km$ in the simulation domain.}
\label{fig:acc}
\end{center}
\end{figure}

In the present 2.5D study we don't study the full feedback cycle (this will be done in a forthcoming 3D study).
However, we find the average accretion rate during the active phase of the jets is about what is required to power the jets.
To eject jets with a total initial kinetic energy of $E_j \simeq 1-5 \times 10^{51} \erg$, a mass of
$\sim M_{\rm acc} \simeq E_j (0.5 G M_{NS}/R_{NS})^{-1} \simeq 0.01-0.05 M_\odot$ should be accreted.
Here we find, as presented in Fig. \ref{fig:acc}, that the total accreted mass is $\sim 0.1 M_\odot$.
Namely, we require that $\sim 10-50 \%$ of the accreted mass after bounce will be through an accretion disk.
The rest of the mass has not enough specific angular momentum to form an accretion disk.
In a forthcoming 3D study the accretion process will be studied in more details.

\section{SUMMARY}
\label{sec:summary}

Different mechanisms for exploding most, or even all, core collapse supernovae
(CCSNe) have been discussed in the literature in recent years.
These include the recently proposed
jittering-jets feedback mechanism (for references see section
\ref{sec:intro}).
This study reports the first hydrodynamical numerical simulations of the
flow structure in the jittering-jets explosion mechanism.
This first study was limited to axisymmetrical flow structure, where a 2D
grid was used to simulate 3D flows, namely, 2.5D simulations.
Two jets, one at each side of the equatorial plane of the numerical
grid, were injected in ten launching episodes chosen
arbitrarily to mimic the stochastic behavior of the jittering-jets
mechanism.
In this grid each jet is a actually a conical shell.
The directions of launching relative to the upper half of the symmetry axis
in the three runs presented here are given in Table \ref{Tab:Table1},
 and the schematic illustration is shown in Fig. \ref{fig:angles}.

Maps of some flow variables of the three different runs are presented in
Figs. \ref{fig:global2sE} - \ref{fig:t2_local}.
From these we learn the following: \\
(1) The more powerful and shorter ($1 \s$) activity duration of the jets
(run 1s20-50) leads to more spherical outflow at early times, and the
inflow region near the equatorial plane is wider, as compared with runs
where the launching activity lasts for two seconds.

(2) Even when jets launching is confined to near the symmetry axis, core
material is efficiently expelled from near the equatorial plane.
A quantitative study of this effect requires 3D simulations. These are
planned in the near future.

(3) Many vortices are formed in the interaction region, inside the bubbles
formed by the shocked jets' material, and outside the bubbles.
These vortices play a role in determining the conditions inside the bubbles,
and hence influence the nuclear reactions inside the bubbles that might form
some r-process elements.
This process will be studied in a future paper.
The vortices are also connected to the inflow of gas: they can be formed in
the shear layer between inflow and outflow, and they can
push material toward the center, hence feeding the newly born neutron star.
The vortices located between inflow and outflow can be clearly seen in Figs.
\ref{fig:zoom} and \ref{fig:t2_local}.

(4) The jittering-jets mechanism is a negative feedback mechanism: The jets expel the
inflowing gas from the core and by that shut themselves out as they prevent
further accretion onto the accretion disk.
The energy required to expel the core is the minimum energy required from
the jets. Since efficiency is not $100 \%$, the typical explosion energy is
few times the binding energy, $\sim 10^{51} \erg$, and above.
In low mass stars the core's binding energy is lower, and weaker explosions
are possible.
Even that feedback is not included in our calculations, in section \ref{sec:acc} we find
that the accreted mass during the jets' activity phase
is sufficient to power the jets under the assumptions of the model. This is
very encouraging for future 3D simulations that will
treat the entire feedback cycle self-consistently.

Overall, we find our results to strengthen the jittering-jets model as
explosion mechanism for {\it all} CCSNe.
To firmly establish the jittering-jets model as the explosion mechanism for
CCSNe, more sophisticated simulations in 3D are required.
These are planned to be conducted in the near future.

\title{Acknowledgments:}
\label{sec:Ack}
{{{  {We thank an anonymous referee for detailed and through report that substantially improved the presentation of our results.} }}}
This research was supported by the Asher Fund for Space Research at the
Technion, the E. and J. Bishop Research Fund
at the Technion, and the USA-Israel Binational Science Foundation.
The {\sc flash} code used in this work is developed in part by the US Department
of Energy under Grant No.
B523820 to the Center for Astrophysical Thermonuclear Flashes at the
University of Chicago.
The simulations were performed on the TAMNUN HPC cluster at the Technion.


\label{lastpage}


\begin{thebibliography}{}\addcontentsline{toc}{section}{References}
\bibitem[Arcavi et al.(2012)]{Arcavi2012} Arcavi, I., Gal-Yam, A.,
Cenko, S.~B., et al.\ 2012, \apjl, 756, L30

\bibitem[Baade \& Zwicky(1934)]{Badde1934} Baade, W., \& Zwicky, F.\ 1934, Physical Review, 46, 76

\bibitem[Bethe \& Wilson(1985)]{bethe1985} Bethe, H.~A., \& Wilson, J.~R.\ 1985, \apj, 295, 14

\bibitem[Bisnovatyi-Kogan et al.(1976)]{Bisnovatyi1976} Bisnovatyi-Kogan, G.~S., Popov, I.~P., \& Samokhin, A.~A.\ 1976, \apss, 41, 287
Zurich style
Sliced ​​veal in mushroom cream sauce , served with hash browns
\bibitem[Blondin \& Mezzacappa(2007)]{BlondinMezzacappa2007} {{{  {Blondin, J.~M., \& Mezzacappa, A.\ 2007, \nat, 445, 58}  }}}

\bibitem[Bruenn et al.(2013)]{Bruenn2013} Bruenn, S.~W., Mezzacappa, A., Hix, W.~R., et al.\ 2013, \apjl, 767, L6

\bibitem[Burrows \& Lattimer(1985)]{Burrows1985} Burrows, A., \& Lattimer, J.~M.\ 1985, \apjl, 299, L19

\bibitem[Burrows 
\& Goshy(1993)]{Burrows1993}{ Burrows, A., \& Goshy, J.\ 1993, \apjl, 416, L75 }

\bibitem[Burrows et al.(1995)]{Burrows1995} Burrows, A., Hayes, J., \& Fryxell, B.~A.\ 1995, \apj, 450, 830

\bibitem[Burrows et al.(2007)]{Burrows2007} Burrows, A., Dessart,
L., Livne, E., Ott, C.~D., \& Murphy, J.\ 2007, \apj, 664, 416

\bibitem[Colgate \& White(1966)]{Colgate1966} Colgate, S.~A., \& White, R.~H.\ 1966, \apj, 143, 626

\bibitem[Couch (2013)]{Couch2013} Couch, S.~M. 2013, Presented in the Fifty-one erg meeting, Raleigh, May 2013.

\bibitem[Couch 
\& O'Connor(2013)]{couch2013arXiv} { Couch, S.~M., \& O'Connor, E.~P.\ 2013, arXiv:1310.5728 }

\bibitem[Couch et al.(2009)]{Couch2009} Couch, S.~M., Wheeler,
J.~C., \& Milosavljevi{\'c}, M.\ 2009, \apj, 696, 953

\bibitem[Couch et al.(2011)]{Couch2011} Couch, S.~M., Pooley, D., Wheeler, J.~C., \& Milosavljevi{\'c}, M.\ 2011, \apj, 727, 104

\bibitem[Dolence et al.(2013)]{Dolence2013} Dolence, J.~C., 
Burrows, A., Murphy, J.~W., \& Nordhaus, J.\ 2013, \apj, 765, 110 

\bibitem[Fern{\'a}ndez(2010)]{Fernandez2010} {{{  {Fern{\'a}ndez, R.\ 2010, \apj, 725, 1563}  }}}

\bibitem[Fryer \& Warren(2002)]{Fryer2002} Fryer, C.~L., \& Warren, M.~S.\ 2002, \apjl, 574, L65

\bibitem[Fryxell et al.(2000)]{Fryxell2000} Fryxell, B., Olson, K.,
Ricker, P., et al.\ 2000, \apjs, 131, 273

\bibitem[Gilkis \& Soker(2013)]{GilkisSoker2013}  {{{ {Gilkis, A. \& Soker, N.\ 2013, work presented at the F.O.E. Fifty-One Erg, Raleigh, NC, , May 2013;
 \url{http://grb.physics.ncsu.edu/FOE2013/WEB/abstracts.html }
}   }}}

\bibitem[Hanke et al.(2012)]{Hanke2012} Hanke, F., Marek, A.,
M{\"u}ller, B., \& Janka, H.-T.\ 2012, \apj, 755, 138

\bibitem[Hanke et al.(2013)]{Hanke2013} Hanke, F., Mueller, B.,
Wongwathanarat, A., Marek, A., \& Janka, H.-T.\ 2013, arXiv:1303.6269

\bibitem[H{\"o}flich et al.(2001)]{Hoflich2001} H{\"o}flich, P., Khokhlov, A., \& Wang, L.\ 2001, 20th Texas Symposium on relativistic astrophysics, 586, 459

\bibitem[Itoh et al.(1996)]{Itoh1996} Itoh, N., Hayashi, H., Nishikawa, A., \& Kohyama, Y.\ 1996, \apjs, 102, 411

\bibitem[Janka(2012)]{Janka2012} Janka, H.-T.\ 2012, Annual
Review of Nuclear and Particle Science, 62, 407

\bibitem[Janka(2013)]{Janka2013} Janka, H.-T.\ 2013, Presented in the Fifty-one erg meeting, Raleigh, May 2013.

\bibitem[Khokhlov et al.(1999)]{Khokhlov1999} Khokhlov, A.~M.,
H{\"o}flich, P.~A., Oran, E.~S., et al.\ 1999, \apjl, 524, L107

\bibitem[Kuroda et al.(2012)]{Kuroda2012} Kuroda, T., Kotake, K.,
\& Takiwaki, T.\ 2012, \apj, 755, 11

\bibitem[Lazzati et al.(2011)]{Lazzati2011} Lazzati, D., Morsony, B.~J., Blackwell, C.~H., \& Begelman, M.~C.\ 2011, arXiv:1111.0970

\bibitem[LeBlanc
\& Wilson(1970)]{LeBlanc1970} LeBlanc, J.~M., \& Wilson, J.~R.\ 1970, \apj, 161, 541

\bibitem[Liebend{\"o}rfer et al.(2005)]{Liebend2005} Liebend{\"o}rfer, M., Rampp, M., Janka, H.-T., \& Mezzacappa, A.\ 2005, \apj, 620, 840

\bibitem[Lopez et al.(2013)]{Lopez2013} Lopez, L.~A.,
Ramirez-Ruiz, E., Castro, D., \& Pearson, S.\ 2013, \apj, 764, 50

\bibitem[MacFadyen et al.(2001)]{MacFadyen2001} MacFadyen, A.~I.,
Woosley, S.~E., \& Heger, A.\ 2001, \apj, 550, 410

\bibitem[Marek
\& Janka(2009)]{Marek2009} Marek, A., \& Janka, H.-T.\ 2009, \apj, 694, 664

\bibitem[Meier et al.(1976)]{Meier1976} Meier, D.~L., Epstein,
R.~I., Arnett, W.~D., \& Schramm, D.~N.\ 1976, \apj, 204, 869

\bibitem[Milisavljevic et al.(2013)]{Milisavljevic2013} Milisavljevic,
D., Soderberg, A., Margutti, R., et al.\ 2013, arXiv:1304.0095

\bibitem[Nordhaus et al.(2010)]{Nordhaus2010} Nordhaus, J., Burrows,
A., Almgren, A., \& Bell, J.\ 2010, \apj, 720, 694

\bibitem[Ott et al.(2008)]{Ott2008} Ott, C.~D., Burrows, A.,
Dessart, L., \& Livne, E.\ 2008, \apj, 685, 1069

\bibitem[Papish
\& Soker(2011)]{papish2011} Papish, O., \& Soker, N.\ 2011, \mnras, 416, 1697

\bibitem[Papish
\& Soker(2012a)]{Papish2012a} Papish, O., \& Soker, N.\ 2012a, Death of Massive Stars: Supernovae and Gamma-Ray Bursts, 279, 377

\bibitem[Papish
\& Soker(2012b)]{Papish2012b} Papish, O., \& Soker, N.\ 2012b, \mnras, 421, 2763

\bibitem[Qian \& Woosley(1996)]{Qian1996} Qian, Y.-Z., \& Woosley, S.~E.\ 1996, \apj, 471, 331

\bibitem[Rantsiou et al.(2011)]{Rantsiouetal2011} {{{ {Rantsiou, E., Burrows, A., Nordhaus, J., \& Almgren, A.\ 2011, \apj, 732, 57}  }}}

\bibitem[Roy et al.(2013)]{Roy2013} Roy, R., Kumar, B., Maund,
J.~R., et al.\ 2013, arXiv:1306.5389

\bibitem[Soker(2010)]{Soker2010} Soker, N.\ 2010, \mnras, 401, 2793

\bibitem[Soker et al.(2013)]{Soker2013} Soker, N., Akashi, M.,
Gilkis, A., et al.\ 2013, Astronomische Nachrichten, 334, 402

\bibitem[Takiwaki et al.(2013)]{Takiwaki2013} Takiwaki, T., Kotake, 
K., \& Suwa, Y.\ 2013, arXiv:1308.5755 

\bibitem[Takaki et al.(2013)]{Takaki2013} Takaki, K., Kawabata,
K.~S., Yamanaka, M., et al.\ 2013, arXiv:1306.5490

\bibitem[Timmes \& Swesty(2000)]{timmes2000} Timmes, F.~X., \& Swesty, F.~D.\ 2000, \apjs, 126, 501

\bibitem[Ugliano et al.(2012)]{Ugliano2012} Ugliano, M., Janka,
H.-T., Marek, A., \& Arcones, A.\ 2012, \apj, 757, 69


\bibitem[Wilson(1985)]{Wilson1985} Wilson, J.~R.\ 1985, Numerical
Astrophysics, 422



\bibitem[Winteler et al.(2012)]{Winteler2012} Winteler, C.,
K{\"a}ppeli, R., Perego, A., et al.\ 2012, \apjl, 750, L22



\bibitem[Woosley
\& Janka(2005)]{Woosley2005} Woosley, S., \& Janka, T.\ 2005, Nature Physics, 1, 147
\end{thebibliography}
\end{document}